\documentclass[10pt,twocolumn,letterpaper]{article}

\usepackage{wacv}
\usepackage{times}
\usepackage{epsfig}
\usepackage{graphicx}
\usepackage{amsmath}
\usepackage{amssymb}
\usepackage{bm}
\usepackage{subfig}
\usepackage{algorithmic}
\usepackage{algorithm}
\usepackage{siunitx}
\usepackage{booktabs}
\usepackage{multirow}
\usepackage{rotating}
\usepackage[table, dvipsnames]{xcolor}
\definecolor{mycolor}{rgb}{0.94, 0.97, 1.0}
\newcommand\figwidth{0.09}
\def\wacvPaperID{1122} % Enter the WACV Paper ID here

\ifwacvfinal
\def\assignedStartPage{1} % *** Enter the assigned starting page number (instead of 9876)
\fi

\ifwacvfinal
\usepackage[breaklinks=true,bookmarks=false]{hyperref}
\else
\usepackage[pagebackref=true,breaklinks=true,colorlinks,bookmarks=false]{hyperref}
\fi

\ifwacvfinal
\else
\fi

\begin{document}

%%%%%%%%% TITLE
\title{Gaze-Guided Class Activation Mapping: Leveraging Human Attention for Network Attention in Chest X-rays Classification}

\author{Hongzhi Zhu\\
University of British Columbia\\
Vancouver, BC, Canada\\
{\tt\small hzhu@ece.ubc.ca}
\and
Septimiu Salcudean\\
University of British Columbia\\
Vancouver, BC, Canada\\
{\tt\small tims@ece.ubc.ca}
\and
Robert Rohling\\
University of British Columbia\\
Vancouver, BC, Canada\\
{\tt\small rohling@ece.ubc.ca}
}

\maketitle
%\thispagestyle{empty}

%%%%%%%%% ABSTRACT
\begin{abstract}
The increased availability and accuracy of eye-gaze tracking technology has sparked  attention-related research in psychology, neuroscience, and, more recently, computer vision and artificial intelligence.
The attention mechanism in artificial neural networks is known to improve learning tasks.
However, no previous research has combined the network attention and human attention.
\par
This paper describes a gaze-guided class activation mapping (GG-CAM) method to directly regulate the formation of network attention based on expert radiologists' visual attention for the chest X-ray pathology classification problem, which remains challenging due to the complex and often nuanced differences among images. GG-CAM is a lightweight ($3$ additional trainable parameters for regulating the learning process) and generic extension that can be easily applied to most classification convolutional neural networks (CNN). GG-CAM-modified CNNs do not require human attention as an input when fully trained.
\par
Comparative experiments suggest that two standard CNNs with the GG-CAM extension achieve significantly greater classification performance. The median area under the curve (AUC) metrics for ResNet50 increases from $0.721$ to $0.776$. For EfficientNetv2 (s), the median AUC increases from $0.723$ to $0.801$. The GG-CAM also brings better interpretability of the network that facilitates the weakly-supervised pathology localization and analysis. \footnote{This manuscript was submitted to WACV 2022 on Aug. 18, 2021.}

\end{abstract}

%%%%%%%%% BODY TEXT
\section{Introduction}
Analogies and comparisons are frequently made between the artificial neural network (ANN) with its biological counterparts.
Multiple explorations and innovations in ANN have been enlightened by concepts from biological neural networks, among which, the attention mechanism, popularized by \cite{vaswani2017attention}, is one of the most prominent research domains \cite{han2020survey}.
The essence of attention (artificially and biologically) is ``the flexible control of limited computational resources", yet its precise mechanism is still pending discovery \cite{hommel2019no,lindsay2020attention}.
\par
Though conceptually intertwined, the attention mechanism applied to computer vision tasks using convolutional neural networks (CNN) does not generally mirror human attention. 
One of the most frequently implemented attention models in CNNs, the multiplicative attention (or the dot-product attention), enables spatial and/or feature-wise attention via explicitly designed network architectures, i.e., the Squeeze-and-Excitation Network \cite{hu2018squeeze} and the Attention Gated Network \cite{schlemper2019attention}. These explicitly crafted architectures can regulate the information flow and automatically form attentions during the training process, but these attentions usually have limited interpretability.
Another branch of research, the class activation mapping (CAM) \cite{zhou2016learning} and its variants \cite{chattopadhay2018grad,selvaraju2017grad,wang2020score}, aims to unveil the specific regions and/or features in the input images that the network is attending to for specific computer vision tasks.
The CAM attention methods can be applied to most generic CNN architectures, i.e., ResNet \cite{he2016deep} and Efficient Net \cite{tan2019efficientnet}, and the resulting attentions are presented as (ideally human interpretable) 2D attention heat maps.
\par
In this paper, unlike previous attempts, we focus on the central question: can human attention be directly exploited to help the formulation of network attention?
\par
To find the answer to this question, we must first quantitatively measure human attention. 
One candidate of such measurement is gaze tracking, which measures the position of one's eye-gaze as a time series \cite{chatelain2018evaluation,zhu2019novel}. 
Due to tracking inaccuracies \cite{zhu2020hand} and the stochastic nature of eye movement \cite{zhu2019neyman}, the raw tracking data are often too noisy to faithfully reflect human attention. Therefore, visual heat maps, generated by 2D clustering and smoothing of the tracked gaze positions, are more commonly employed to represent the distribution of one's visual attention \cite{holmqvist2011eye}.
\par
With the visual heat maps, we propose the method of {\em gaze-guided class activation mapping (GG-CAM)} that uses the visual attention heat map to supervise the formation of the CAM attention in generic CNN architectures for image classification. 
The proposed method, to the best of our knowledge, is the first that combines network attention with human attention. 
When applied to the disease classification tasks for chest X-ray (CXR) images in a public dataset \cite{karargyris2021creation}, GG-CAM displays three major advantages:
\begin{enumerate}
    \item It is a light weight (only $3$ additional trainable parameters) and generic method that can be easily added to most classification CNNs for the integration of human attention to network attention.
    \item The performance of CNN classifications with GG-CAM can be significantly improved, and the fully trained networks no longer require visual attention as reference.
    \item The interpretability of the CNNs is improved, as it can spatially relate pathology to organ location. 
\end{enumerate}
\par
The rest of the paper is organized as follows. 
We first present the background of this paper in Section \ref{sec:Background}, where we review recent advancements on disease classification on CXR images with CNN and deep learning methods using visual attention. We also summarize the CAM method in Section \ref{sec:Background}. Next, In Section \ref{sec:GG-CAM: Proposed Approach}, we propose the GG-CAM method. To validate our method, experiments were conducted, and results are discussed in Section \ref{sec:Experiments}. Lastly, we conclude our paper in Section \ref{sec:Conclusion}.

\section{Background}\label{sec:Background}
\subsection{Deep learning with Chest X-Ray (CXR) images}
CXR imaging is one of the most frequently used medical diagnostic tools, capable of identifying multiple pathologies, i.e., pneumonia, triage, pneumothorax, tuberculosis, COVID-19, and cardiomegaly \cite{ccalli2021deep}. There exist many public CXR datasets, such as ChestX-ray14 \cite{wang2017chestx}, CheXpert \cite{irvin2019chexpert}, MIMIC-CXR \cite{johnson2019mimic} and Ped-Pneumonia \cite{kermany2018large}, that focus on single or multiple pathologies.
\par
For datasets focusing on a single pathology, existing CNNs for classification (ResNet \cite{he2016deep}, MobileNet \cite{howard2017mobilenets}, etc.) can yield outstanding classification performance \cite{khan2021intelligent}. 
For example, in the pediatric pneumonia CXR dataset (Ped-Pneumonia \cite{kermany2018large}), \cite{hashmi2020efficient} reports a binary classification accuracy of $98.4\%$. However, datasets targeting multiple pathologies are more difficult for the CNNs to classify \cite{ccalli2021deep,khan2021intelligent}.
One example is the MIMIC-CXR dataset \cite{johnson2019mimic}, where there are at least $70$ different kinds of pathologies and abnormalities, multiple of which may coexist in a single CXR image \cite{wu2020comparison}. The state-of-the-art CNN classifier for the MIMIC-CXR dataset \cite{johnson2019mimic} is achieved by \cite{wu2020comparison} and \cite{wong2020robust}, which share the same CNN architecture but focus on two different classification tasks. In \cite{wong2020robust}, focus is on the binary classification task (normal versus pathological/abnormal CXR images), and the reported AUC is $0.824$. In \cite{wu2020comparison}, $70$ pathologies and abnormalities are considered for classification, and the AUC for each class ranged from $0.628$ to $0.985$.
\par
Other than the challenges arising from the complexity of CXR images, the fact that most CXR datasets are imbalanced adds an additional level of difficulty for network training and fair benchmarking \cite{khan2021intelligent}. To solve the problem, in \cite{karargyris2021creation}, a balanced dataset derived from the MIMIC-CXR dataset \cite{johnson2019mimic} was proposed. Images in the balanced dataset can be classified into three mutually exclusive categories: normal, pneumonia and cardiomegaly (enlarged heart).
%Additionally, sonographer's gaze movement was collected during the diagnostic process, which is of great use in our research. 
It is also a modality rich dataset that can facilitate with cross-modal learning and multi-task learning researches.

\subsection{Visual heat maps assisted convolutional deep learning}
\label{sec:Visual heat maps assisted convolutional deep learning}
Visual heat maps, as a measurement of one's visual attention, are employed to assist with CNN training and optimization through two primary approaches. 
The most straightforward approach is to feed visual heat maps into the CNN together with the images, as in \cite{cai2018multi,cai2018sonoeyenet,liu2020using,sharma2021multi}, because gaze patterns are task specific \cite{karessli2017gaze}.
By doing so, the network can further process the information conveyed in the visual heat maps to enhance performance; however, this adds an input to the network, making it harder to deploy to real-world tasks.
The other approach avoids the dependency on the visual heat maps through representation learning, by using the visual heat maps to facilitate the learning of representative and robust features through transfer learning \cite{droste2019ultrasound,droste2020discovering,zhang2020human} or multi-task learning \cite{bera2021gaze}. Still, transfer learning methods may suffer from negative transfer or overfitting, and multi-task learning methods commonly introduce a large number additional parameters to the network.
\par
Although existing CNN methods with visual heat maps yield improved performance, the networks are designed to processes the visual heat maps in the same manner as other information sources, leaving the embedded attention characteristics underexplored.

\subsection{Class Activation Mapping (CAM)}
\begin{figure}[t]
    \centering
    \subfloat[Computational blocks for generic classification CNNs]{
	   \includegraphics[width=0.45\textwidth]{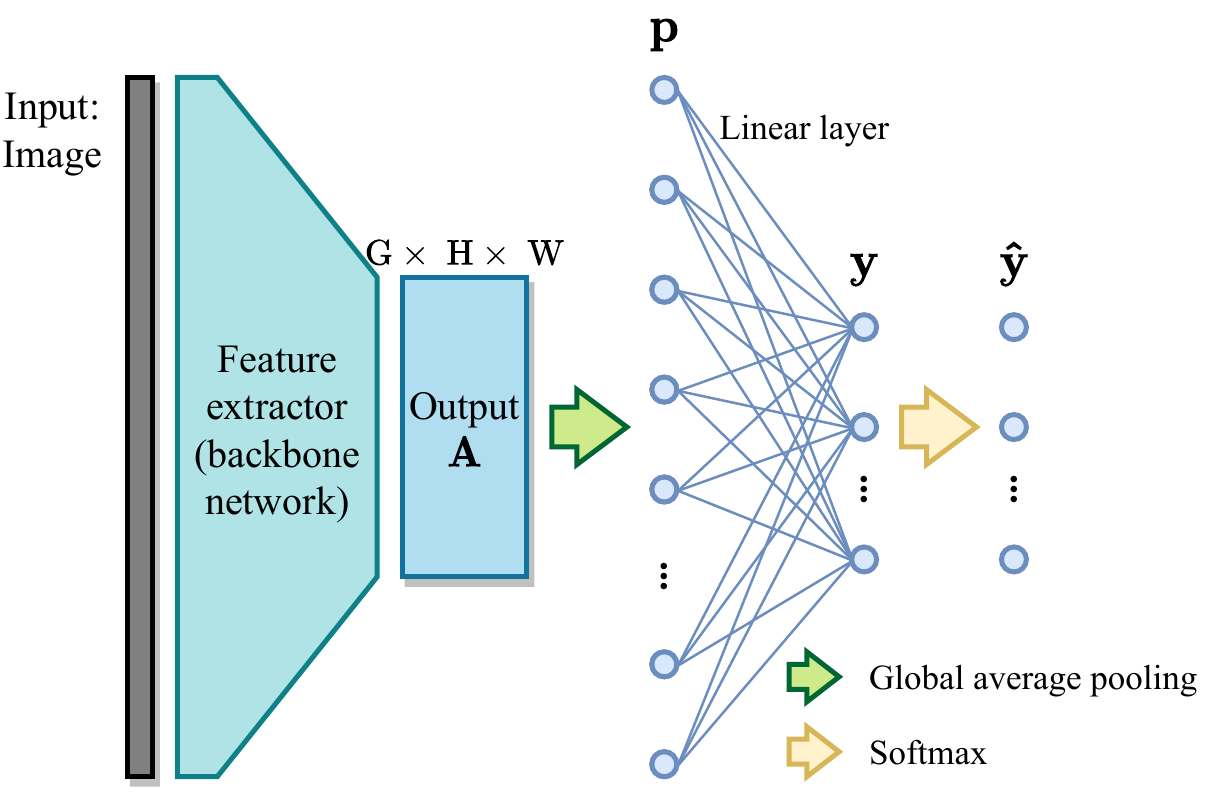}
	   \label{fig:Generic classification CNNs}
	   }
     \vfill
    \subfloat[Computational blocks for CAM-CNN]{
	   \includegraphics[width=0.45\textwidth]{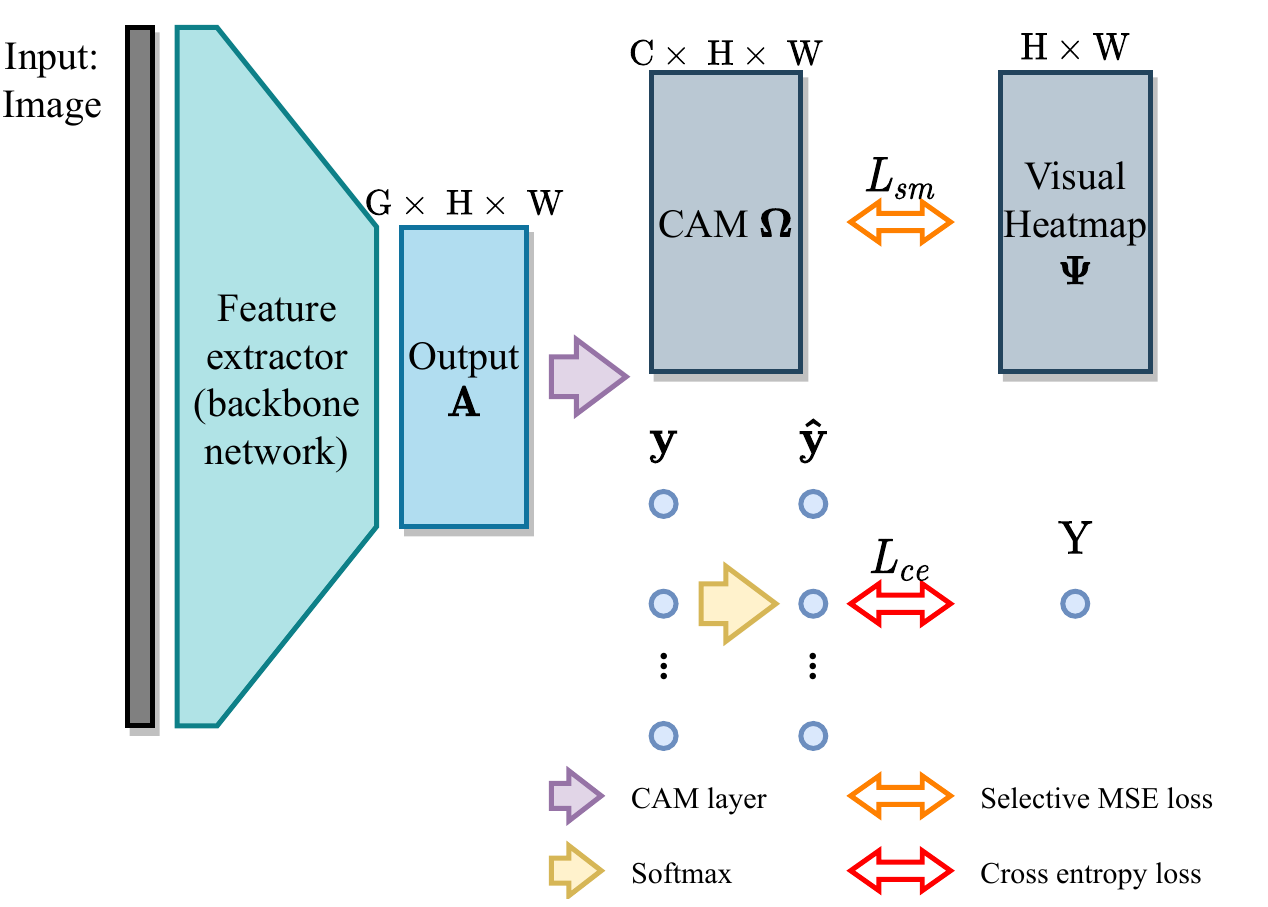}
	   \label{fig:CAM-CNN}
	   }
\caption{Comparison between generic classification CNN and CAM-CNN.}
\label{fig:demographic}
\end{figure}

Most classification CNNs are built with the same consecutive computational blocks (shown in Figure \ref{fig:Generic classification CNNs}): a feature extractor (or called backbone network), a global average pooling layer, a linear layer (or called dense layer or fully connection layer), and a softmax layer (or other normalization operation). To demystify the internal mechanisms of the CNN ``black boxes", the CAM method \cite{zhou2016learning} was proposed. In \cite{zhou2016learning}, they focused on the global average pooling layer and the linear layer in the CNN, from which they extracted 2D attention maps, depicting regions on the input that the network is attending for the outputs. The detailed CAM method is explained next. 
\par
Let $\bm{A}\in\mathbb{R}^{G\times H\times W}$ be the output tensor from the feature extractor, where $G$ stands for the number of features, $H$ and $W$ represent the spatial dimensions; $\bm{p}\in\mathbb{R}^{G}$ be the output tensor from the global average pooling layer; and $\bm{y}\in\mathbb{R}^{C}$ be the output tensor from the linear layer, where $C$ is the number of classes that the network is trying to categorize. Without loss of generality, we neglect the batch dimension for tensors in our paper. Therefore, we have:
\begin{equation}
    \bm{y} = \bm{\Lambda}\bm{p}+\bm{\lambda}
\label{eq:y}
\end{equation}
where $\bm{\Lambda}\in\mathbb{R}^{C\times G}$ and $\bm{\lambda}\in\mathbb{R}^{C}$ are trainable weights and biases, respectively, in the linear layer. More specifically, the $c^{th}$ element in vector $\bm{y}$, $\bm{y}_c$, is related to $A$ through the following equation:
\begin{equation}
\begin{split}
    \bm{y}_c &= \sum_{k=1}^G\left[\bm{\Lambda}_{c,k}\cdot\bm{p}_k\right]+\bm{\lambda}_c\\
    &=\sum_{k=1}^G\left[\bm{\Lambda}_{c,k}\cdot\left(\frac{1}{HW}\sum_{i=1}^H\sum_{j=1}^W\bm{A}_{k,i,j}\right)\right]+\bm{\lambda}_c\\
\end{split}
\label{eq:y_g}
\end{equation}
where $\bm{A}_{k,i,j}$ denotes the element positioned at $(k,i,j)$ in $\bm{A}$. 
Through altering the order of summation in Equation (\ref{eq:y_g}), we have:
\begin{equation}
\bm{y}_c = \frac{1}{HW}\sum_{i=1}^H\sum_{j=1}^W\left[\sum_{k=1}^G\bm{\Lambda}_{c,k}\cdot\bm{A}_{k,i,j}\right]+\bm{\lambda}_c.
\label{eq:y_g2}
\end{equation}
The CAM for the network, $\bm{\Omega}\in\mathbb{R}^{C\times H\times W}$, is defined as:
\begin{equation}
    \bm\Omega_{c,i,j} = \sum_{k=1}^G\bm{\Lambda}_{c,k}\cdot\bm{A}_{k,i,j}
\label{eq:omega}
\end{equation}
which is the inner most summation in Equation (\ref{eq:y_g2}). 
\par
Through Equation (\ref{eq:omega}), we know that $\bm\Omega$ has three dimensions. Let $\bm{\Omega}^c\in\mathbb{R}^{H\times W}$ be the $c^{th}$ slice of $\bm\Omega$ in the first dimension; 
$\bm{\Omega}^c$ is a 2D attention heat map that explains which regions on the input contribute to the decision of predicting class $c$ as the CNN output. This can be better explained by combining Equations (\ref{eq:y_g2}) and (\ref{eq:omega}):
\begin{equation}
    \bm{y}_c = \bar{\bm\Omega}^c+\bm\lambda_c
    \label{eq:y_g3}
\end{equation}
where
\begin{equation}
    \bar{\bm\Omega}^c = \frac{1}{HW}\sum_{i=1}^H\sum_{j=1}^W\bm\Omega_{i,j}^c
    \label{eq:bar_omega}
\end{equation}
is the mean of all elements in $\bm\Omega_{i,j}^c$.
From (\ref{eq:y_g3}) we can see that the value of $\bm{y}_c$ can only be increased if elements in $\bm{\Omega}^c$ are increased. $\bm\Omega$ has been frequently applied for network explanation and weakly unsupervised localization \cite{bae2020rethinking,liang2020weakly}.

\section{GG-CAM: Proposed Approach}\label{sec:GG-CAM: Proposed Approach}
The proposed GG-CAM method has four major components. Firstly, we propose a novel layer, termed CAM layer, to substitute the standard classification head (the global average pooling layer and the linear layer) in a classification CNN. Secondly, we describe the method to generate a visual heat map from raw gaze coordinates. Thirdly, we introduce a novel loss function for supervising network attention with human attention. The last is the multi-task training methods that we adopt to balance the attention supervision and classification tasks. 
\subsection{CAM layer}
From Equations (\ref{eq:y}) through (\ref{eq:omega}), we know that $\bm\Omega$ is embedded in the network though it is not explicitly computed. To better utilize $\bm\Omega$, we propose the CAM layer to explicitly compute $\bm\Omega$ in the network. The CAM layer is created to replace the standard classification head, the global average pooling layer and the linear layer, in a generic classification CNN as shown in Figure \ref{fig:Generic classification CNNs}. The CAM layer has the same number of trainable parameters, $\bm\Lambda$ and $\bm\lambda$, as in a linear layer. Mathematically, the CAM layer takes $\bm{A}$ as input, uses Equation (\ref{eq:omega}) to compute $\bm\Omega$, and then applies (\ref{eq:y_g3}) to $\bm\Omega$ for $\bm{y}$. 
Therefore, in principle, the only difference between using the standard classification head and using the CAM layer as the classification head is the rearrangement of the interchangeable mathematical operations. 
By doing so, as depicted in Figure \ref{fig:CAM-CNN}, there are two possible outputs from the CAM layer: $\bm{y}$ and $\bm\Omega$, which facilitates the integration of human attention into the network.

\subsection{Visual heat map generation}
Let $\bm{E}_{1:M}=\{\bm{E}_1,...,\bm{E}_M\}$ be a sequence of 2D coordinates of tracked eye-gaze position at time indices $1$ through $M$ when viewing image $\bm{I}\in\mathbb{R}^{H_0\times W_0}$. Let $\mathcal{H}(\cdot)$ be the 2D histogram function that takes a sequence of coordinates as input and outputs a 2D heat map. One of the multiple possible steps to generate the visual heat map, $\bm\Psi$, is  summarized in Algorithm \ref{alg:1}.

\begin{algorithm}
\caption{Generate visual heat map $\bm\Psi$}
\label{alg:1}
\begin{algorithmic}[1]
\STATE Construct $\mathcal{H}$ with horizontal and vertical bins corresponding to the positions of each pixel in $\bm{I}$.
\STATE {$\bm\Psi\gets \mathcal{H}(\bm{E}_{1:M})$}, where $\bm\Psi\in\mathbb{R}^{H_0\times W_0}$.
\STATE {$\bm\Psi\gets \mathcal{G}(\bm\Psi;B)$}, where $\mathcal{G}(\cdot;B)$ is the Gaussian blur operation with standard deviation parameter $B$ for the Gaussian kernel \cite{frederick1998efficient}.
\STATE {$\bm\Psi\gets \text{Resample}(\bm\Psi)$}, such that the new $\bm\Psi$ has dimension $H\times W$.
\STATE {$\bm\Psi\gets \text{Normalize}(\bm\Psi)$}, such that elements in the new $\bm\Psi$ lies between $[0,1]$
\RETURN {$\bm\Psi$}
\end{algorithmic}
\end{algorithm}

\subsection{Attention supervision}
Usually, to train a classification CNN, the cross entropy loss, $\mathcal{L}_{ce}$, is used:
\begin{equation}
    \mathcal{L}_{ce} = -\ln{\hat{\bm{y}}_{Y}}
    \label{eq:L_ce}
\end{equation}
where $Y\in\{1,2,...,C\}$ is the true class label, and $\hat{\bm{y}}=\text{Softmax}(\bm{y})$. 
Substituting (\ref{eq:y_g3}) in (\ref{eq:L_ce}) yields:
\begin{equation}
    \mathcal{L}_{ce} = -\bar{\bm{\Omega}}^Y-\bm{\lambda}_Y+\ln\left[\sum_{i=1}^C\exp\left(\bar{\bm{\Omega}}^i+\bm{\lambda}_i\right)\right].
    \label{eq:L_ce2}
\end{equation}
From (\ref{eq:L_ce2}), we see that the cross entropy loss only focuses on the mean values of each $\bm{\Omega}^i$, such that different $\bm{\Omega}$ can possibly result in the same $\mathcal{L}_{ce}$ as long as the mean value for each $\bm{\Omega}^i$, $i\in\{1,2,...,C\}$, is unchanged. 
This opens the way for us to regulate $\bm{\Omega}$ without jeopardizing the network's performances.
\par
To regulate network attention with human attention, we proposed the selective mean square error (MSE) loss, $\mathcal{L}_{sm}$, which supervises $\bm\Omega$ with $\bm\Psi$ during training:
\begin{equation}
    \mathcal{L}_{sm} = \frac{1}{HW}\sum_{i=1}^H\sum_{j=1}^W\left(\tilde{\bm\Omega}^Y_{i,j}-\bm{\Psi}_{i,j}\right)^2
    \label{eq:L_sm}
\end{equation}
where $Y\in\{1,2,...,C\}$ is the true class label for input $\bm{I}$. 
$\tilde{\bm\Omega}^c$ is related to ${\bm\Omega}^c$ via the following equation:
\begin{equation}
    \tilde{\bm\Omega}^c = \text{Sigmoid}(\alpha{\bm\Omega}^c)
    \label{eq:tilde_omage}
\end{equation}
where $c = 1,2,...,C$ and $\alpha>0$ is a trainable scalar.
In Equation (\ref{eq:L_sm}), we can see that for input $\bm{I}$ with label $Y$, only $\bm{\Omega}^Y$ is supervised, and other $\bm{\Omega}^i$, $i\neq Y$, are neglected. That is based on the fact that one's visual pattern is task specific as described in Section \ref{sec:Visual heat maps assisted convolutional deep learning}. We expect that the biological and artificial attention for a heart condition is different from that for a lung infection. Therefore, visual attention for input $\bm{I}$ with label $Y$ should only be used to supervise $\bm{\Omega}^Y$.
\par
\begin{figure}[t]
    \centering
    \subfloat[EfficientNetv2]{
	   \includegraphics[width=0.225\textwidth]{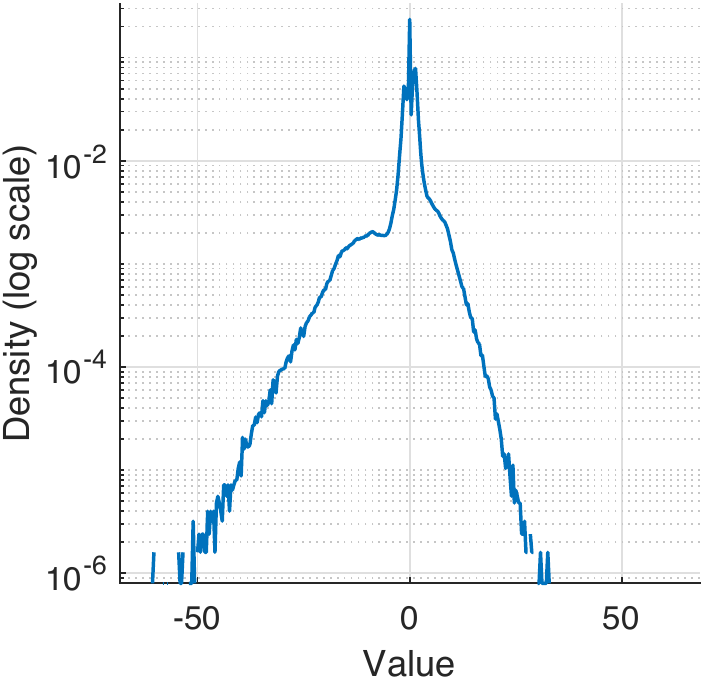}
	   \label{fig:ENet-REF-CAM-elements}
	   }
     \hfill
    \subfloat[ResNet]{
	   \includegraphics[width=0.225\textwidth]{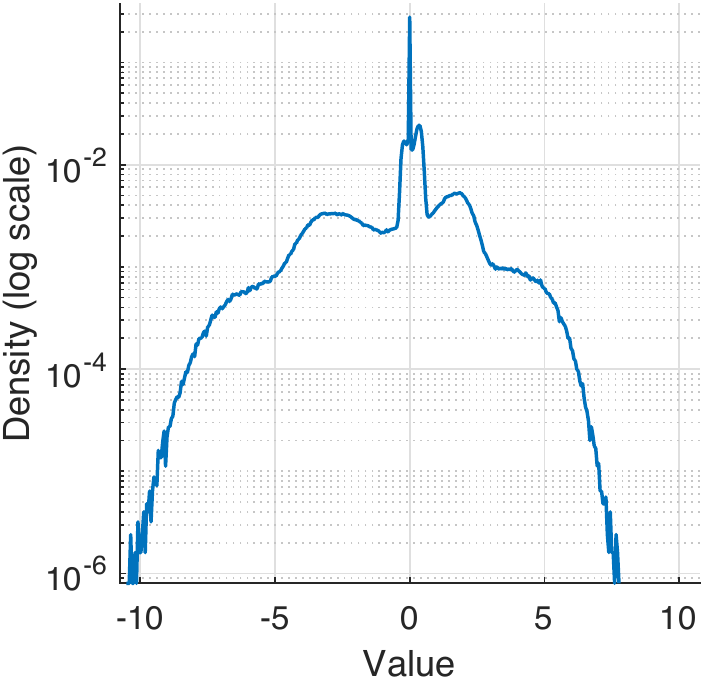}
	   \label{fig:RNet-REF-CAM-elements}
	   }
\caption{Distribution of CAM elements.}
\label{fig:CAM distribution}
\end{figure}

Equation (\ref{eq:tilde_omage}) is applied to elements in $\bm\Omega$ for two main reasons. The first reason lies in the boundlessness of $\bm\Omega$: elements in $\bm\Omega$ can have any values. Figure \ref{fig:CAM distribution} shows the distribution of elements in $\bm{\Omega}$ for two fully-trained CNNs, ResNet and EfficientNetv2, for the CXR classification task. The distribution functions are shown in log-scale. We can see that both distributions are heavy-tailed with median, mode and mean values around $0$. However, elements in the visual heat maps, according to Algorithm \ref{alg:1}, are bounded between $0$ and $1$. Hence, to scale $\bm\Omega$, the Sigmoid function is used. Additionally, we introduce an additional trainable parameter $\alpha$ in Equation (\ref{eq:tilde_omage}) to allow for more flexibility in the supervision.

\subsection{Balanced training}\label{sec:Balanced training}
In the previous section, other than the classification loss $\mathcal{L}_{ce}$, we introduce $\mathcal{L}_{sm}$ for attention supervision. With two losses for a single CNN, the multi-task learning methods should be adopted to balance the effects from each loss functions. 
In this paper, we use the following equation to perform multi-task learning \cite{liebel2018auxiliary}:
\begin{equation}
    \mathcal{L} = \frac{1}{2\sigma_{sm}^2}\mathcal{L}_{sm}+\frac{1}{\sigma_{ce}^2}\mathcal{L}_{ce}+\ln(\sigma_{sm}+1)+\ln(\sigma_{ce}+1)
    \label{eq:L}
\end{equation}
where $\sigma_{sm}>0$, $\sigma_{ce}>0$ are trainable parameters that can dynamically weigh the two losses $\mathcal{L}_{sm}$ and $\mathcal{L}_{sm}$; and $\ln(\sigma_{sm}+1)$ and $\ln(\sigma_{ce}+1)$ are used to penalize large values of $\sigma_{sm}$ and $\sigma_{ce}$, respectively. Equation (\ref{eq:L}) was derived from \cite{kendall2018multi}, where $\sigma_{sm}$ and $\sigma_{ce}$ are used to measure the uncertainties of the corresponding loss terms. The advantage of Equation (\ref{eq:L}) as compared to the loss used in \cite{kendall2018multi} is that the loss function is non-negative and can not diverge to minus infinity during training.
\par
Additionally, experimental results suggest the initialization of parameters $\sigma_{sm}$ and $\sigma_{ce}$ effects the performance of the network. As classification is the primary task for the network, $\mathcal{L}_{ce}$ should be emphasised, especially towards the ending phase of the training process. Therefore, we initialize $\sigma_{sm}$ with a small value close to $0$, and set $\sigma_{ce}=1$. A small initial $\sigma_{sm}$ will force the network to focus more on reducing $\mathcal{L}_{sm}$ at the starting phase of the training process. Gradually, $\sigma_{sm}$ increases, and the network no longer places its primary focus on $\mathcal{L}_{sm}$. In this way, $\mathcal{L}_{sm}$ and $\mathcal{L}_{ce}$ have similar contributions to $\mathcal{L}$, especially towards the end of the training process.

\section{Experiments}\label{sec:Experiments}
To validate our method, we apply GG-CAM to two standard classification CNNs: ResNet50 \cite{he2016deep} and EfficientNetv2 (s) \cite{tan2021efficientnetv2}. To present the comparative analysis, we firstly introduce the dataset that we use for this research, as well as settings during the training process. Then, we present quantitative results on the network's classification performance and interpretability. Lastly, limitations and future improvements are discussed.

\subsection{Dataset}
We use the multi-modal CXR dataset from \cite{karargyris2021creation} (available at PhysioNet \cite{goldberger2000physiobank}). It contains $1083$ CXR images originated from the CXR-MIMIC dataset \cite{johnson2019mimic}. Accompanying each image is a label (either normal, cardiomegaly or pneumonia), segmentation (mediastanum, aortic knob, left and right lungs), radiology report (text and audio), and tracked eye-gaze coordinates from an expert radiologist examining the image. The GP3 Gaze Tracker (by Gazepoint) with \SI{60}{\Hz} sampling frequency was used, and the tracked gaze positions have an accuracy around \SI{1}{\degree} of visual angle (approximately $100$-$400$ pixels on the image depending on the screen size, eye-to-screen distance and calibration accuracy). It is a balanced dataset with even share for each labels.
\par
For the training, validation, and testing, we use $70\%$, $10\%$, and $20\%$ of the dataset, respectively. Each sub-dataset is randomly generated and preserves the label balancing. 
The raw CXR images in the dataset follow the digital imaging and communications in medicine (DICOM) standard and have very high resolution, i.e., $2544\times3056$ pixels. 
Before feeding the images to CNNs, we downsample the images so that the heights and weights are halved. To reduce the high dynamic range of DICOM images, we normalize all images to the range of $[0,1]$.

\subsection{Training}
We use the PyTorch framework for the implementation, training, and testing of CNNs. During the training process, we implement the ``reduce learning rate on plateau'' mechanism, such that the learning rate will be reduced to $10\%$ once the validation loss has not decreased for $P$ consecutive epochs. For all networks, we optimize the following hyper-parameters: learning rate, optimizer, and $P$. The optimizers we use are: Adam, Adamax, stochastic gradient descent (SGD), and SGD+momentum, with their default settings. For GG-CAM modified networks, we also optimize for the initialization of $\sigma_{sm}$, and the Gaussian blur parameter $B$.
Experimental results indicate that the network's performance is not sensitive to $B$ when $B$ is in range $[200,1000]$.
Learning rate and the initialization of $\sigma_{sm}$ play more important roles in the learning process.
\par
We use ``EffNet'' and ``ResNet'' to abbreviate EfficientNetv2 (s) and ResNet50, respectively. We append ``GG-CAM'' to the GG-CAM modified network name (connected by a plus sign) to differentiate standard CNNs from GG-CAM modified CNNs, respectively, i.e., EffNet and ResNet+GG-CAM.
Due to the randomness in the network's performance even with identical hyper-parameters, all results for CNNs shown in this paper are based on $5$ independent trainings with same hyper-parameters.
In Table \ref{tab:hyper}, we summarize the optimized hyper-parameters for each CNN. 
Adam is the best optimizer for all CNNs, and thus not included in the table.

\begin{table}[t]
\centering
\tabcolsep=0.12cm
\begin{tabular}{@{}ccccc@{}}
\toprule
Network         & learning rate & $P$  & $\sigma_{sm}$  & $B$                \\ \midrule
EffNet+GG-CAM & $6.0\times10^{-3}$          & $40$       & $1.4\times10^{-11}$  & $600$\\
ResNet+GG-CAM & $7.0\times10^{-3}$          & $30$       & $2.0\times10^{-9}$   & $500$\\
EffNet & $1.0\times10^{-3}$          & $25$       & -     & -                      \\
ResNet & $1.0\times10^{-4}$          & $50$       & -     & -                      \\
 \bottomrule
\end{tabular}
\caption{Optimized hyper-parameters for CNNs. $P$ is the patience parameter for reducing the learning rate on the plateau. $B$ is the Gaussian blur parameter.}
\label{tab:hyper}
\end{table}

\begin{figure}[t]
    \centering
    \subfloat[AUC for EffNets]{\includegraphics[width = 0.225\textwidth]{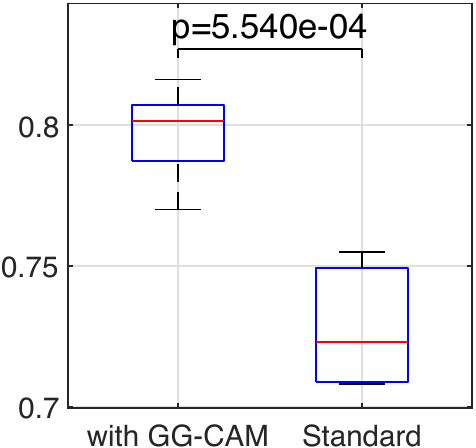}
    \label{fig:AUC_ENet}}
    \hfill
    \subfloat[AUC for ResNets]{\includegraphics[width = 0.225\textwidth]{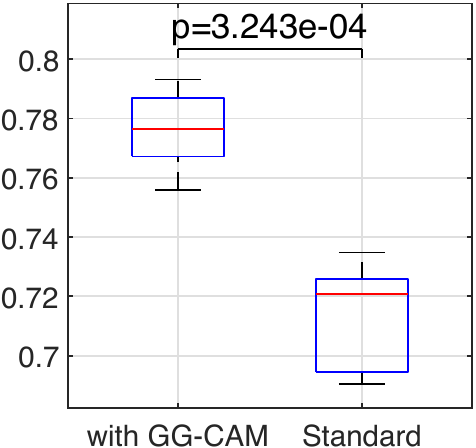}
    \label{fig:AUC_RNet}}
    \vfill
    \subfloat[Class cardiomegaly interpretability for EffNets]{\includegraphics[width = 0.225\textwidth]{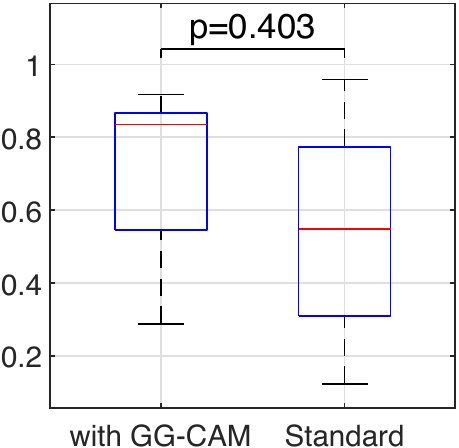}
    \label{fig:heart_ENet}}
    \hfill
    \subfloat[Class cardiomegaly interpretability for ResNets]{\includegraphics[width = 0.225\textwidth]{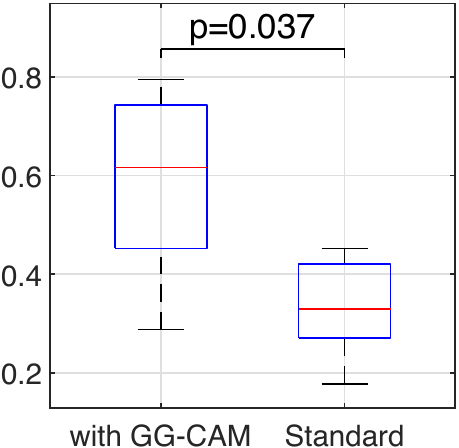}
    \label{fig:heart_RNet}}
    \vfill
    \subfloat[Class pneumonia interpretability for EffNets]{\includegraphics[width = 0.225\textwidth]{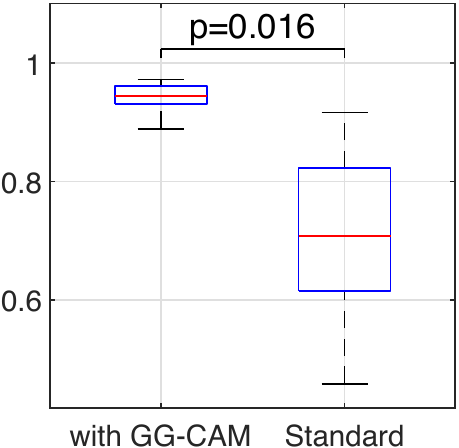}
    \label{fig:lung_ENet}}
    \hfill
    \subfloat[Class pneumonia interpretability for ResNets]{\includegraphics[width = 0.225\textwidth]{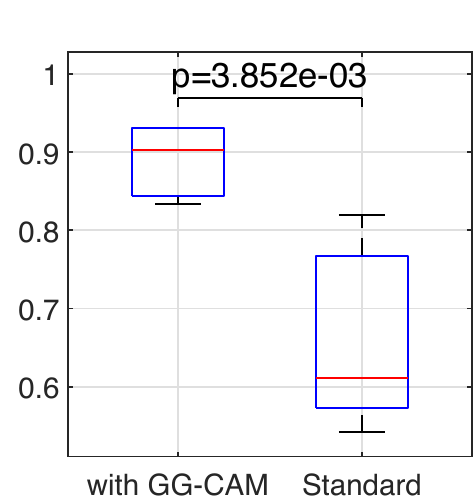}
    \label{fig:lung_RNet}}
    \caption{Results with ANOVA statistical test}
    \label{fig:hand eye coordination}
\end{figure}

\begin{table*}[t]
    \centering
    \tabcolsep=0.06cm
    \begin{tabular}{c|c|c|c|c|c|c}
    \toprule
         Class & \multicolumn{2}{c|}{normal ($Y=1$)} & \multicolumn{2}{c|}{cardiomegaly ($Y=2$)} & \multicolumn{2}{c}{pneumonia ($Y=3$)}\\ \cmidrule(lr){1-7}
    Metrics & precision & recall & precision & recall & precision & recall \\ \midrule
    \rowcolor{mycolor}
        EffNet+GG-CAM & $\bm{0.649 \pm 0.042}$ & $\bm{0.611 \pm 0.080}$ & $\bm{0.660 \pm 0.015}$ & $0.699 \pm 0.136$ & $\bm{0.500 \pm 0.042}$ & $0.486 \pm 0.086$ \\ 
        ResNet+GG-CAM & $0.633 \pm 0.024$ & $0.556 \pm 0.040$ & $0.658 \pm 0.023$ & $0.699 \pm 0.054$ & $0.450 \pm 0.026$ & $\bm{0.500 \pm 0.054}$ \\ 
        
    \rowcolor{mycolor}
        EffNet & $0.569 \pm 0.035$ & $0.611 \pm 0.147$ & $0.654 \pm 0.065$ & $0.699 \pm 0.053$ & $0.418 \pm 0.065$ & $0.389 \pm 0.178$ \\ 
        ResNet & $0.586 \pm 0.043$ & $0.528 \pm 0.153$ & $0.540 \pm 0.048$ & $\bm{0.712 \pm 0.228}$ & $0.375 \pm 0.028$ & $0.389 \pm 0.122$ \\ \bottomrule
    \end{tabular}
    \caption{Detailed metrics for each class and CNN. Values are presented as median statistics followed by the standard deviation after the $\pm$ sign. The best metrics are highlighted in bold font.}
    \label{tab:metrics}
\end{table*}

\begin{table*}[t]
\centering
\begin{tabular}{@{}cccc@{}}
\toprule
$\bm\Omega^Y$                     & $Y=1$ (normal)               & $Y=2$ (cardiomegaly) & $Y=3$ (pneumonia) \\ \midrule
\begin{turn}{90} $\quad\bm\Psi\quad$  \end{turn}
&
\includegraphics[width = \figwidth\textwidth]{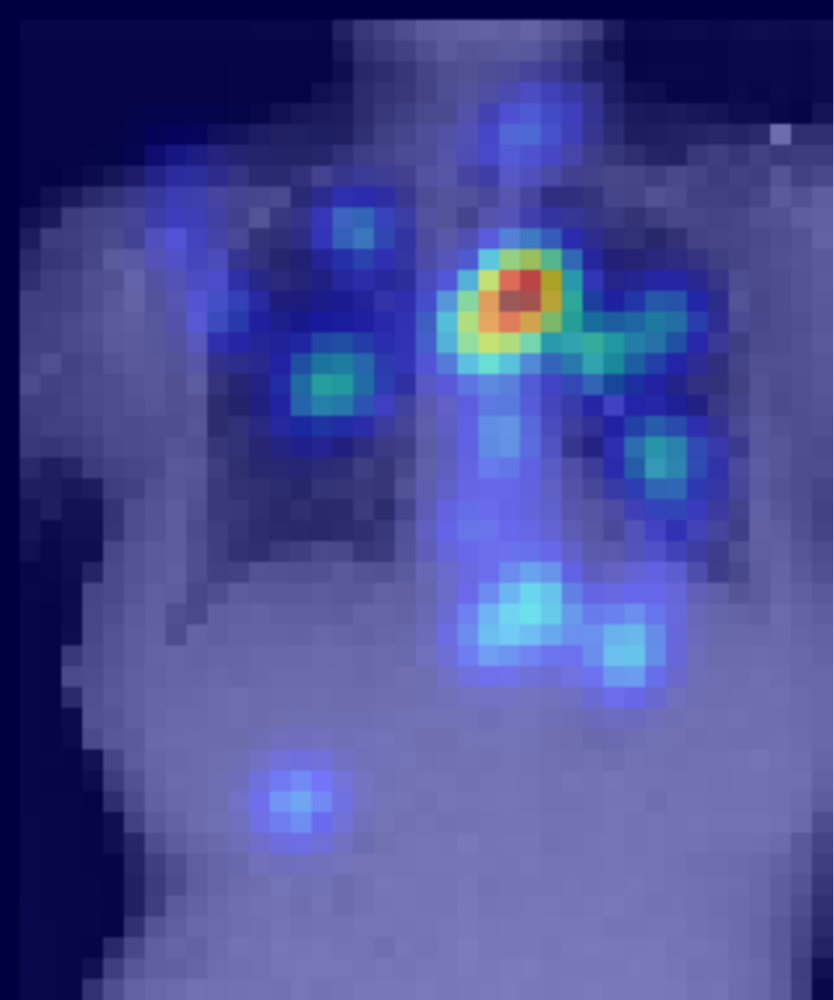}
\includegraphics[width = \figwidth\textwidth]{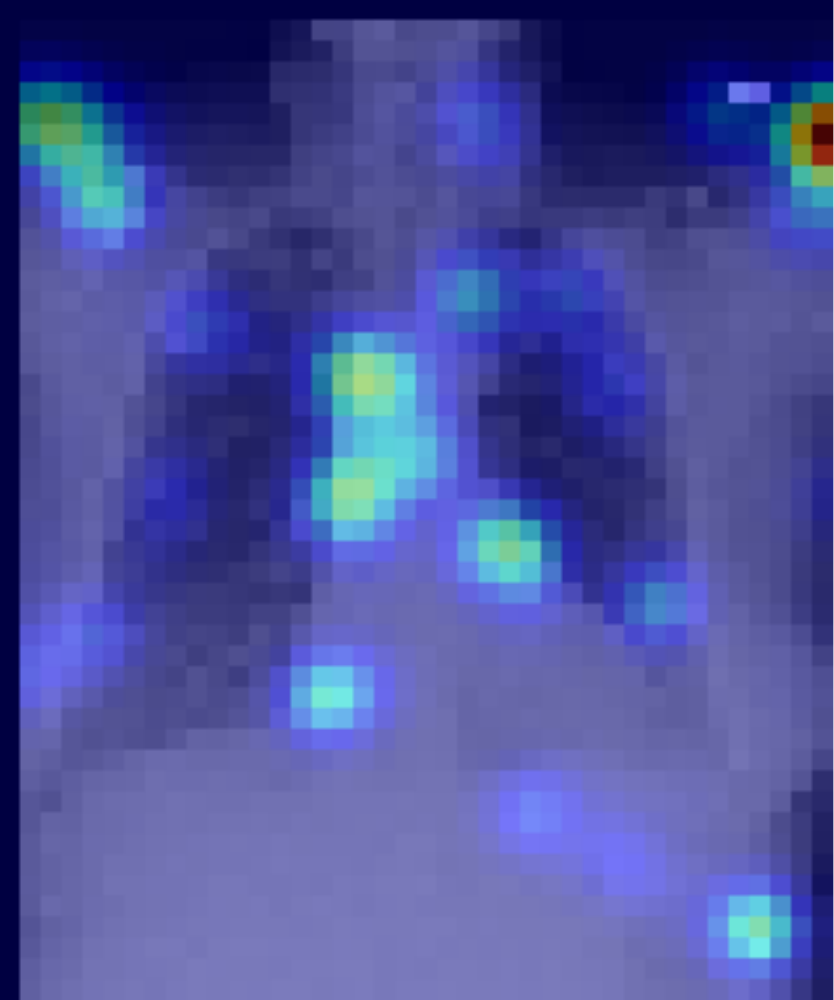}
\includegraphics[width = \figwidth\textwidth]{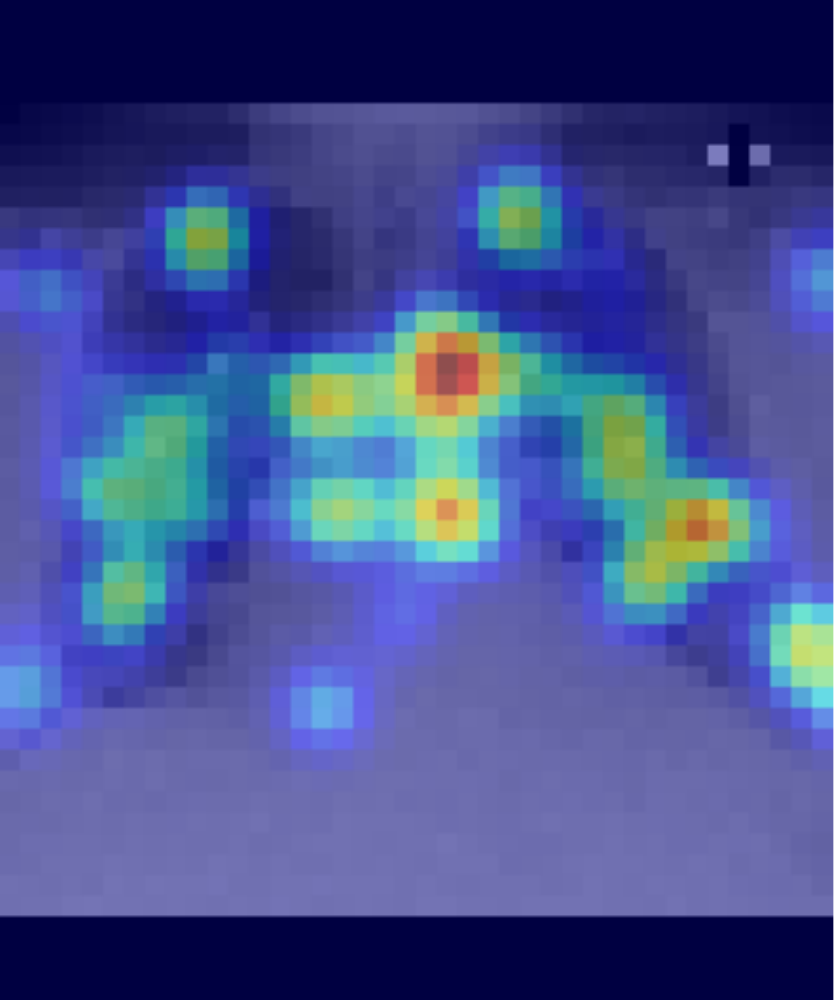}
&  
\includegraphics[width = \figwidth\textwidth]{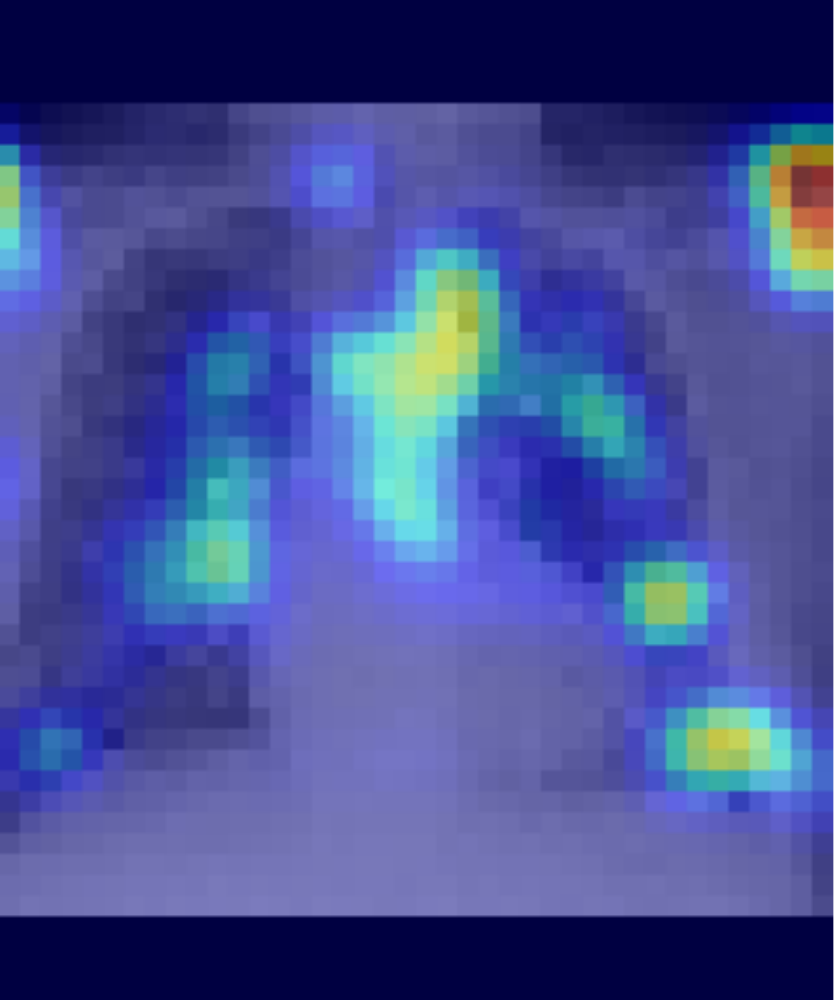}
\includegraphics[width = \figwidth\textwidth]{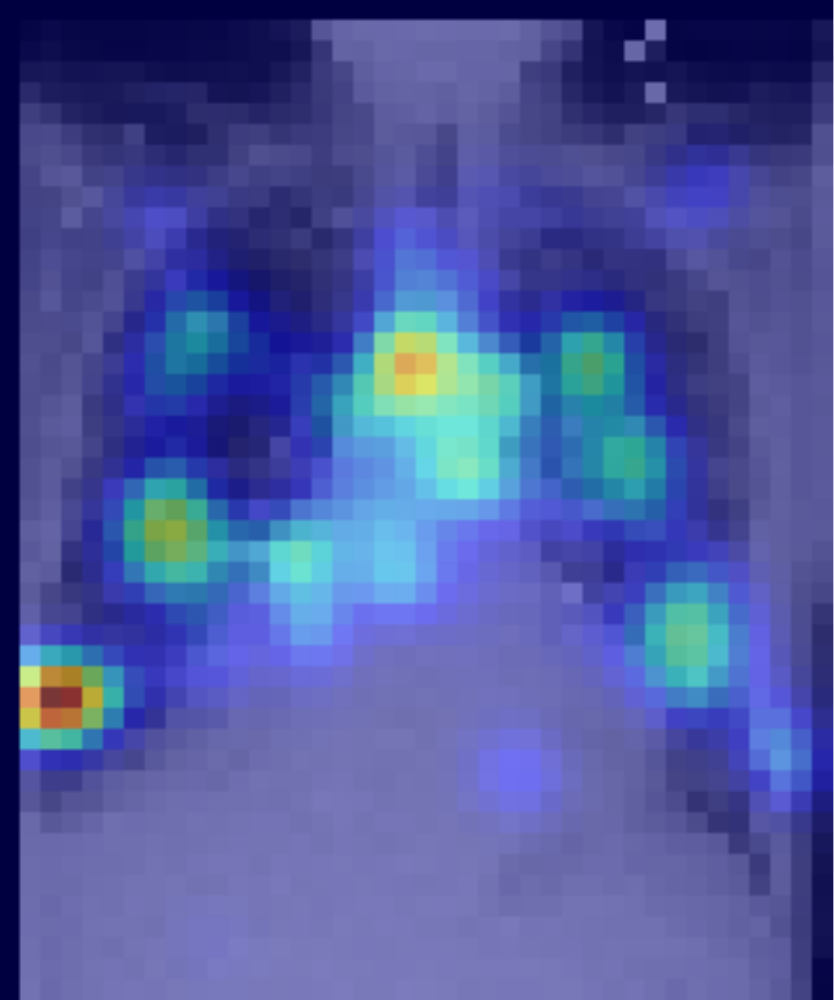}
\includegraphics[width = \figwidth\textwidth]{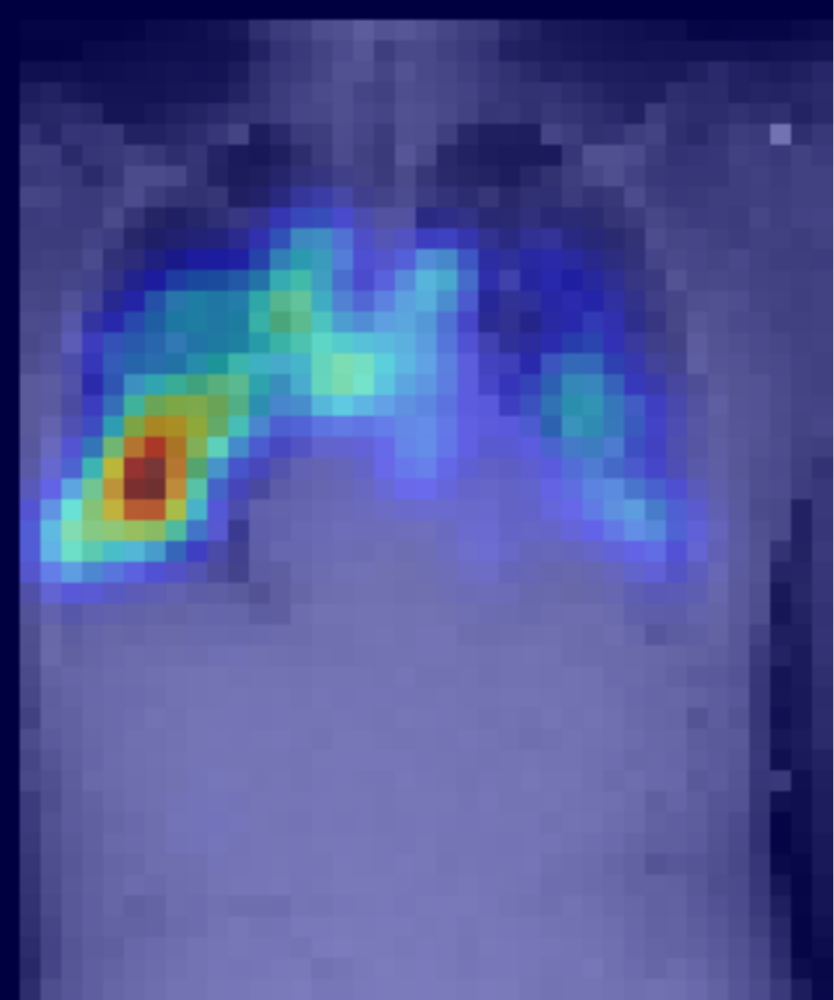}
&
\includegraphics[width = \figwidth\textwidth]{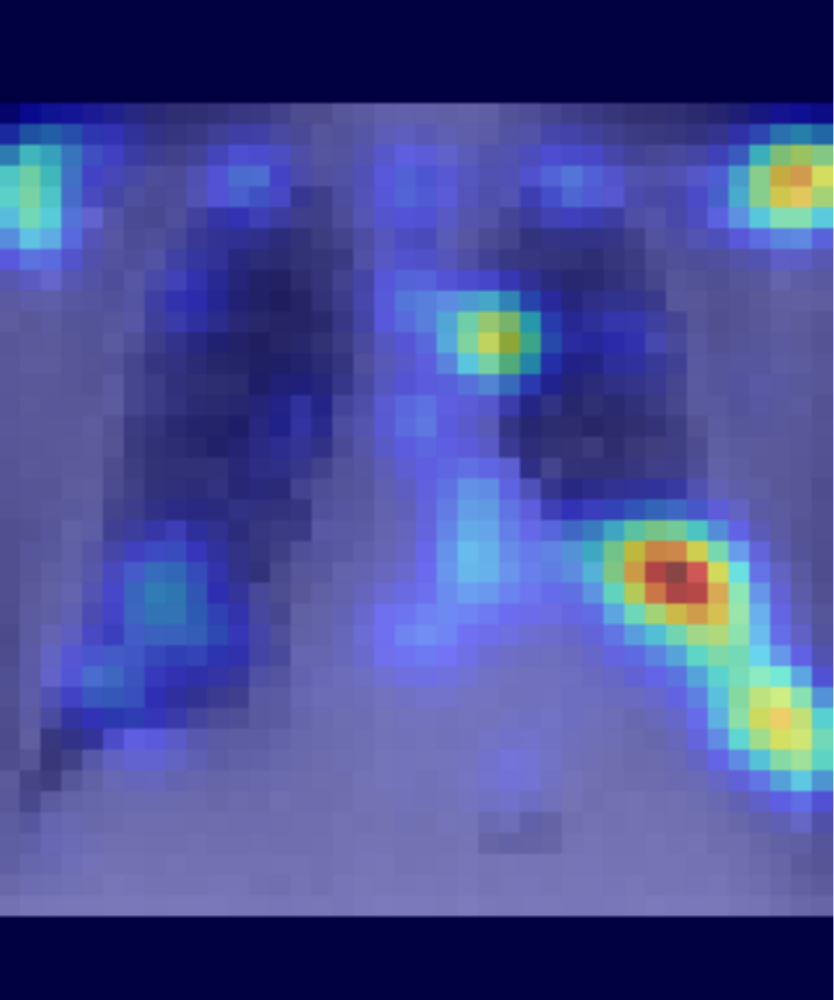}
\includegraphics[width = \figwidth\textwidth]{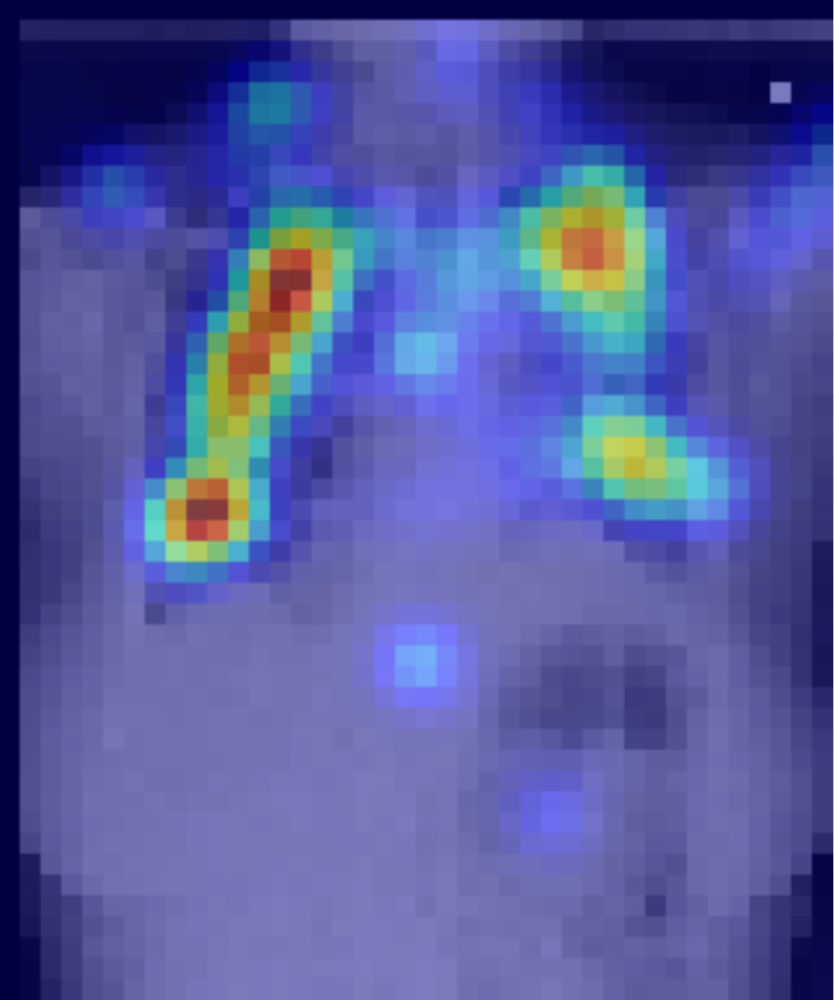}
\includegraphics[width = \figwidth\textwidth]{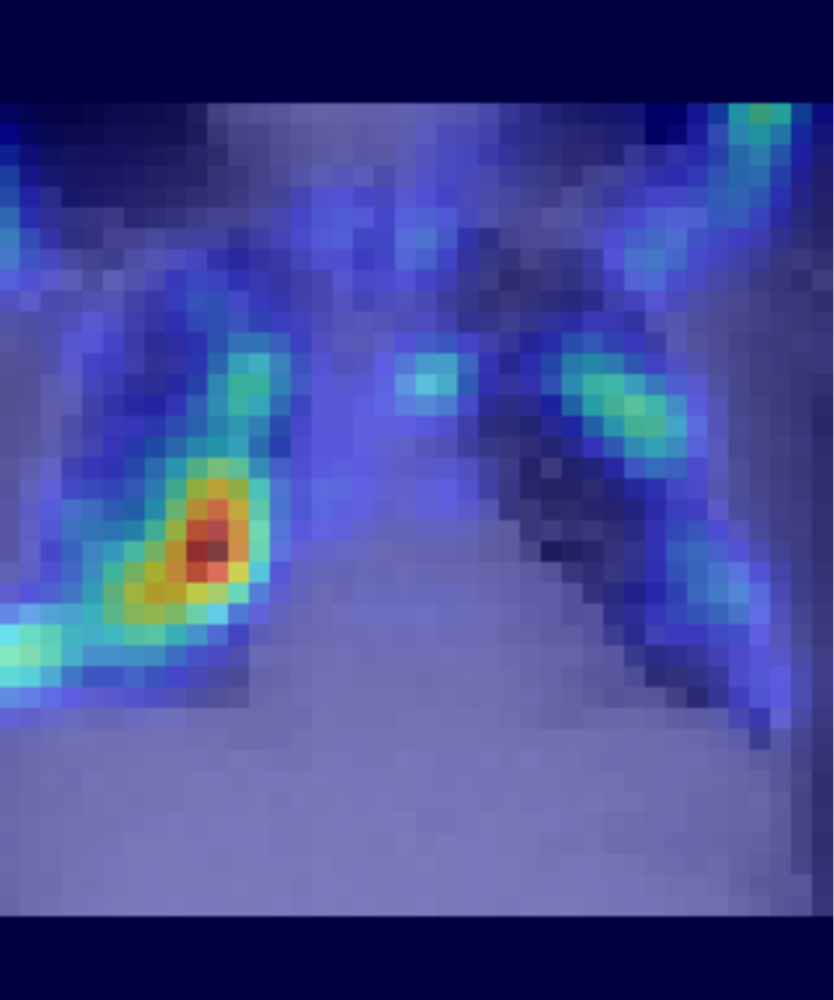}
\\
\begin{turn}{90} \parbox{2cm}{EffNet\\+GG-CAM}  \end{turn} 
&
\includegraphics[width = \figwidth\textwidth]{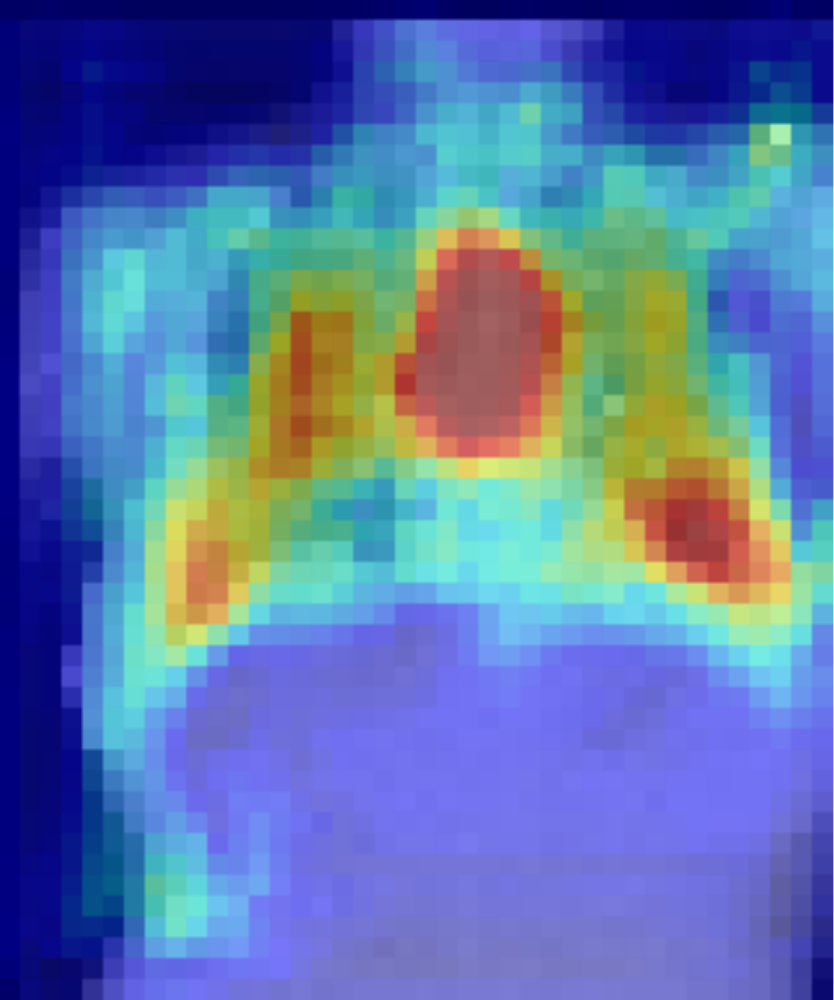}
\includegraphics[width = \figwidth\textwidth]{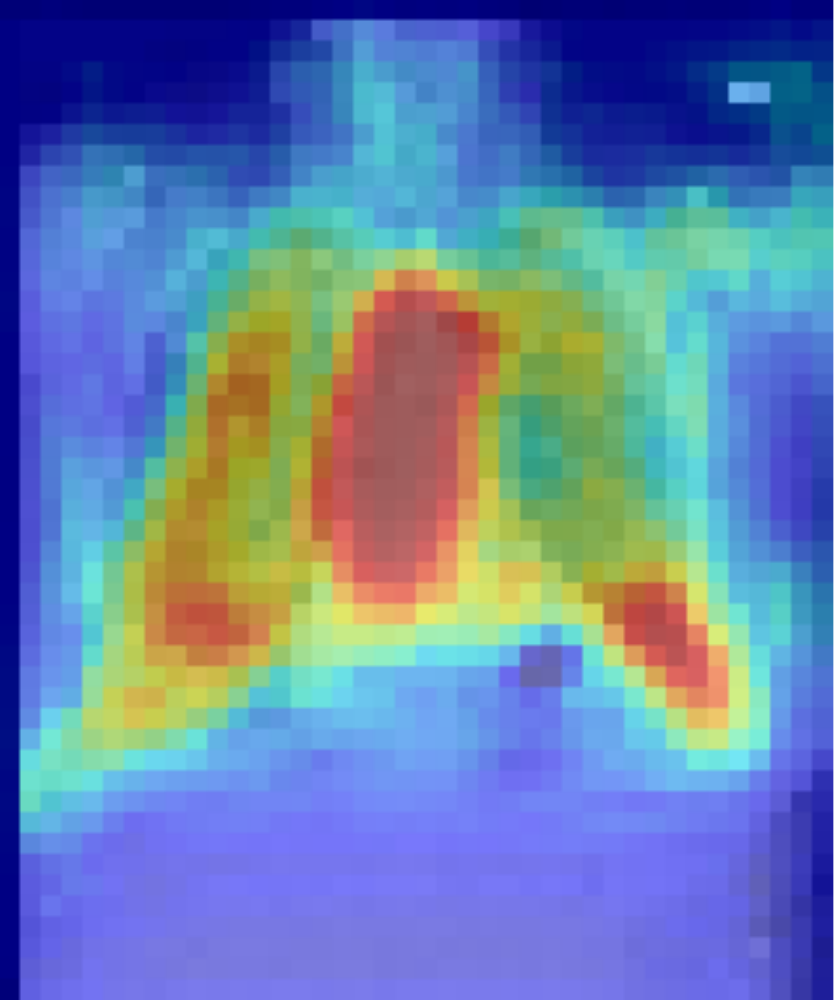}
\includegraphics[width = \figwidth\textwidth]{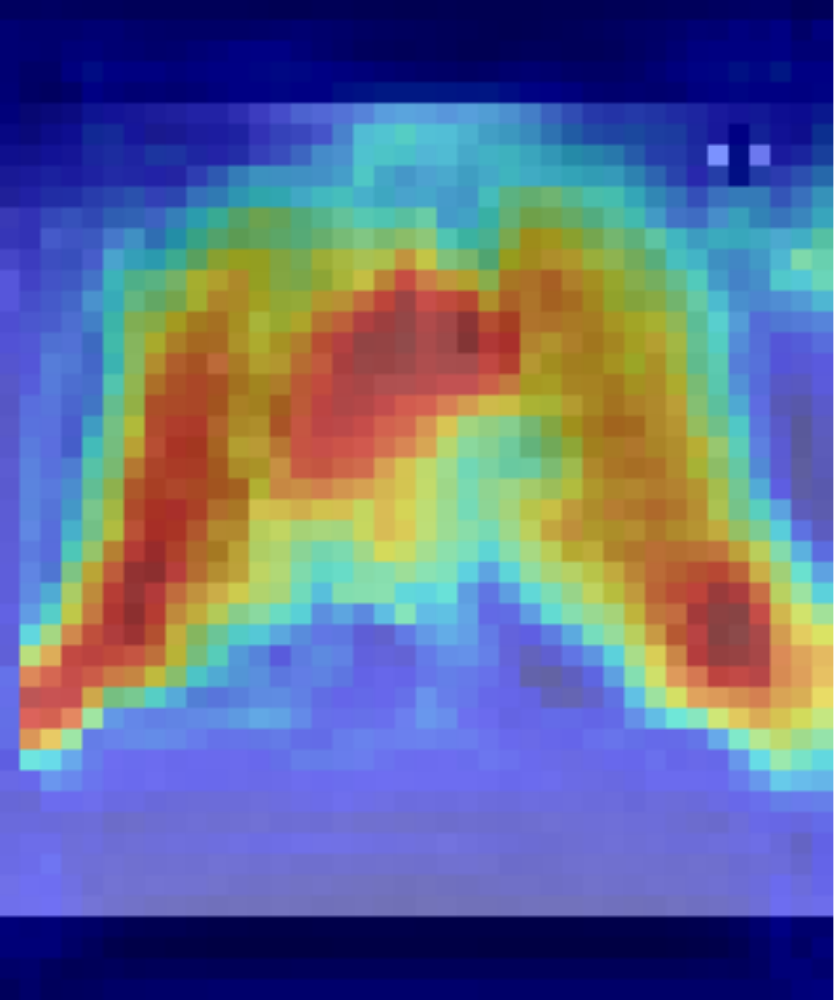}
& 
\includegraphics[width = \figwidth\textwidth]{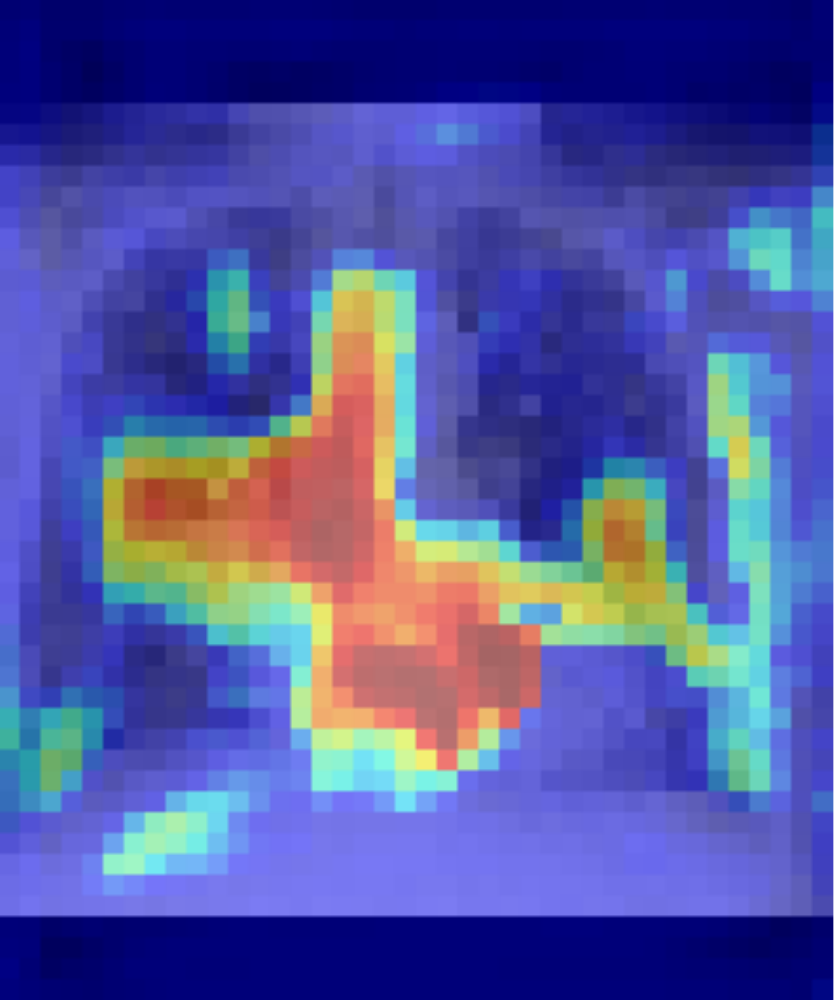}
\includegraphics[width = \figwidth\textwidth]{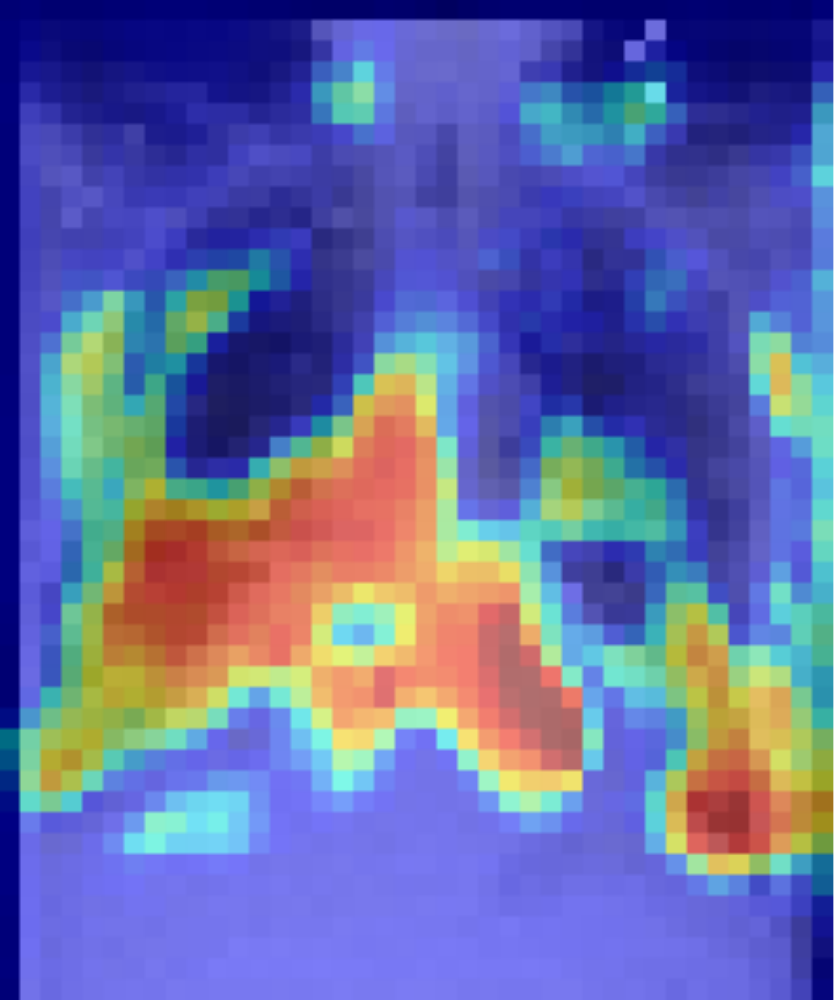}
\includegraphics[width = \figwidth\textwidth]{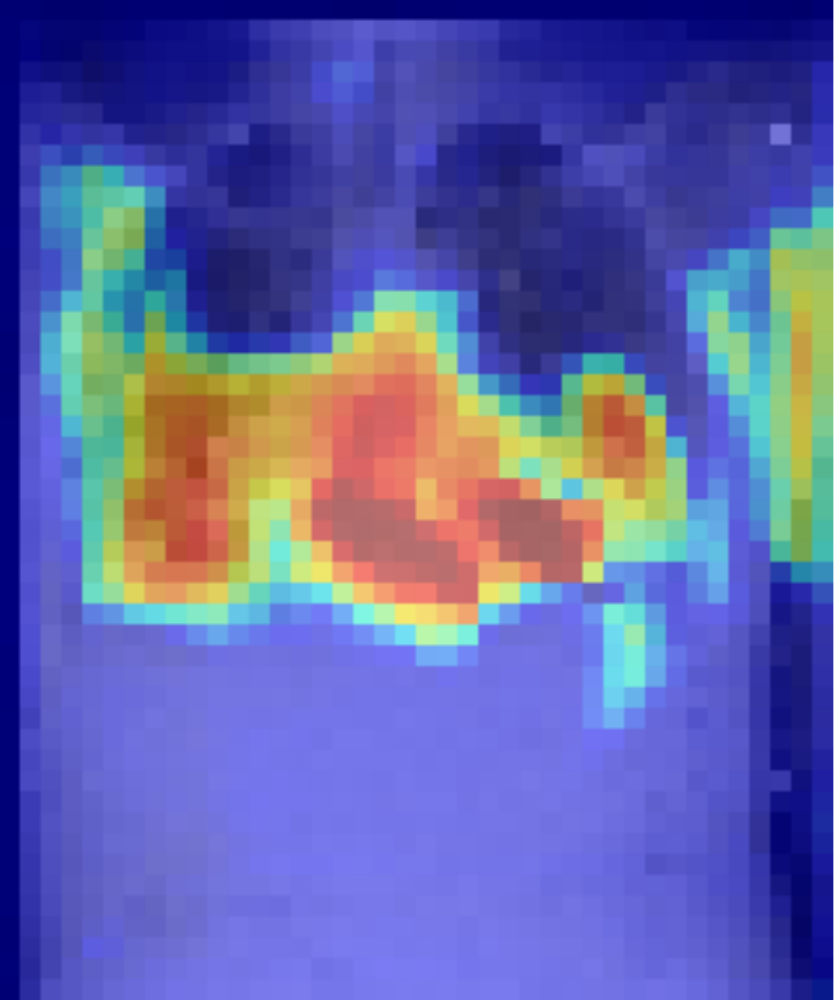}
& 
\includegraphics[width = \figwidth\textwidth]{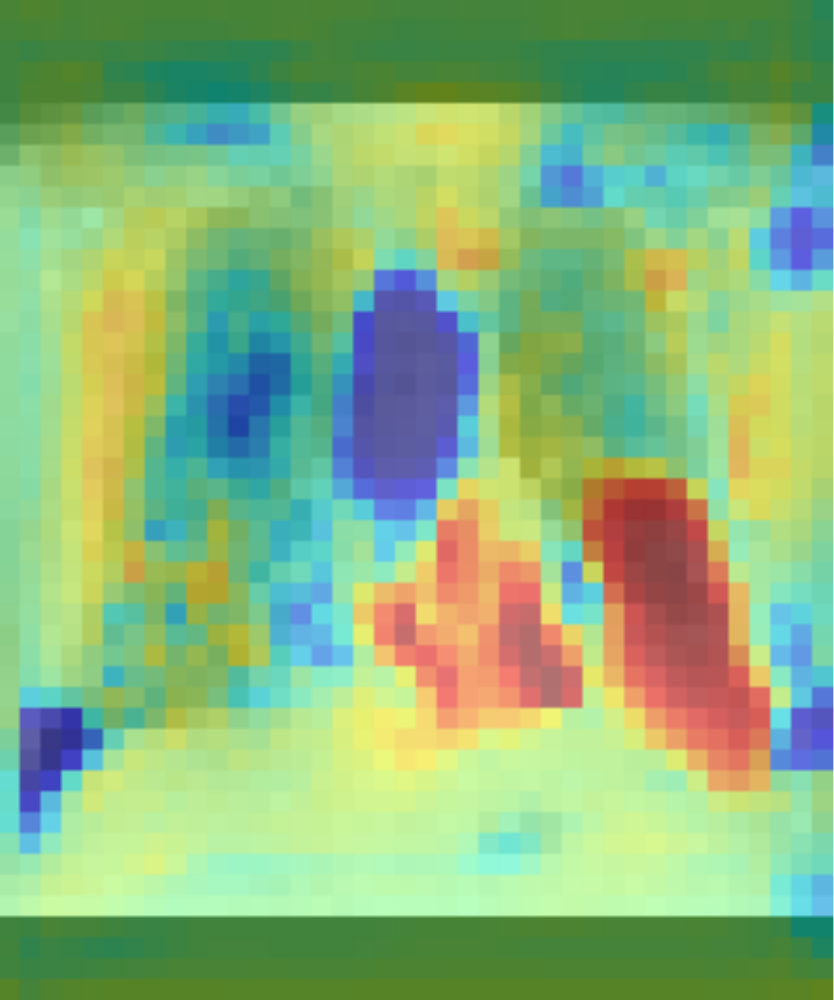}
\includegraphics[width = \figwidth\textwidth]{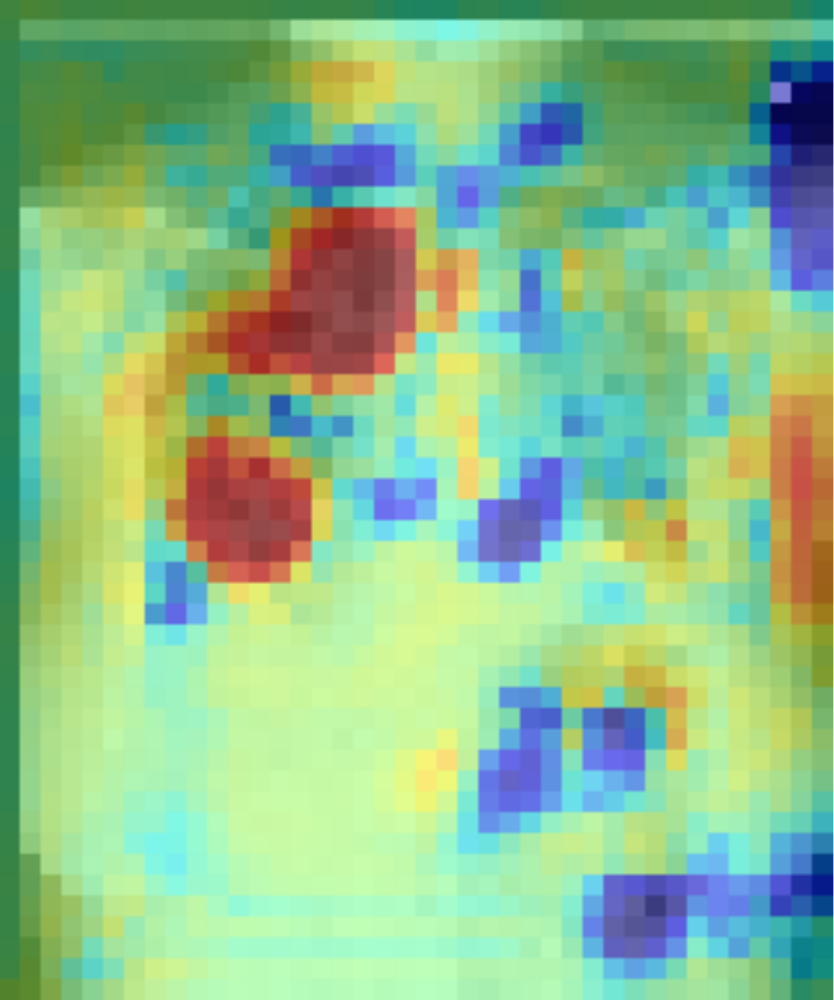}
\includegraphics[width = \figwidth\textwidth]{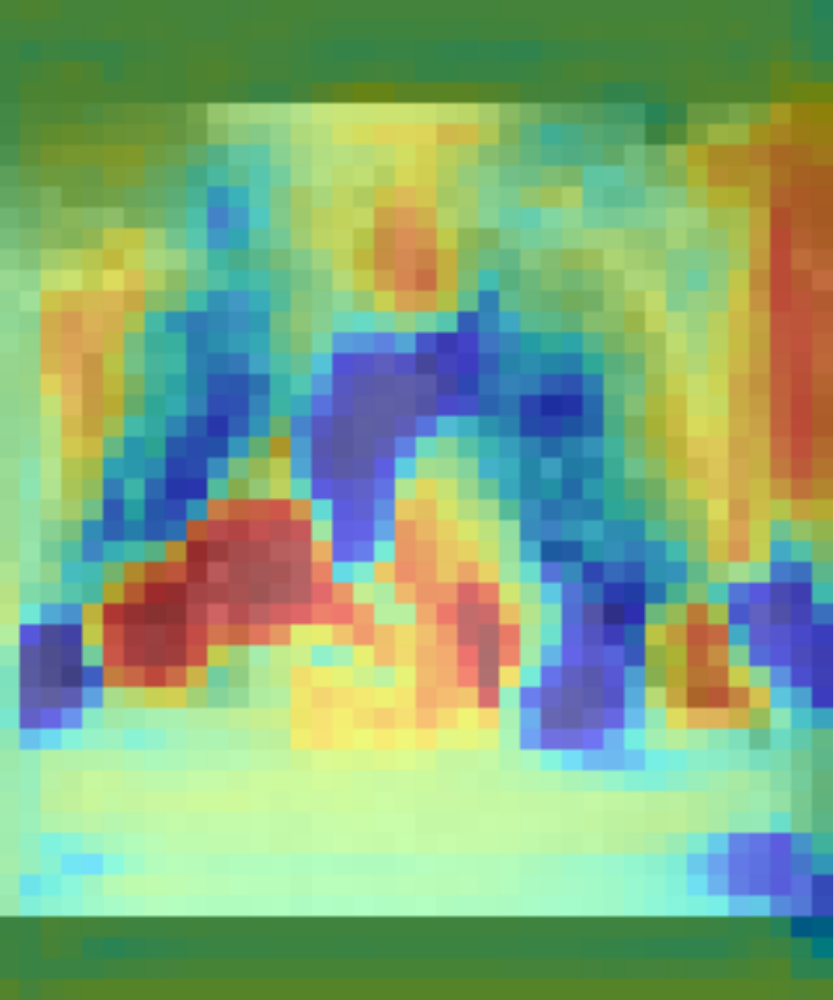}
\\
\begin{turn}{90} \parbox{2cm}{ResNet\\+GG-CAM}  \end{turn}
&
\includegraphics[width = \figwidth\textwidth]{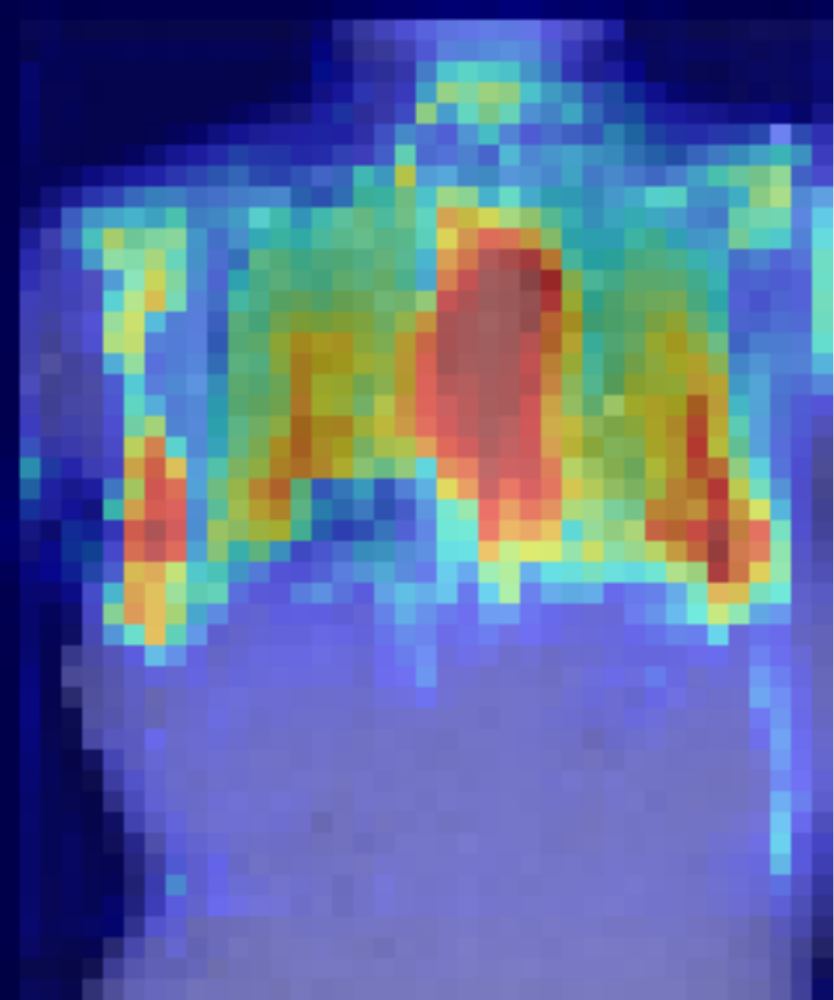}
\includegraphics[width = \figwidth\textwidth]{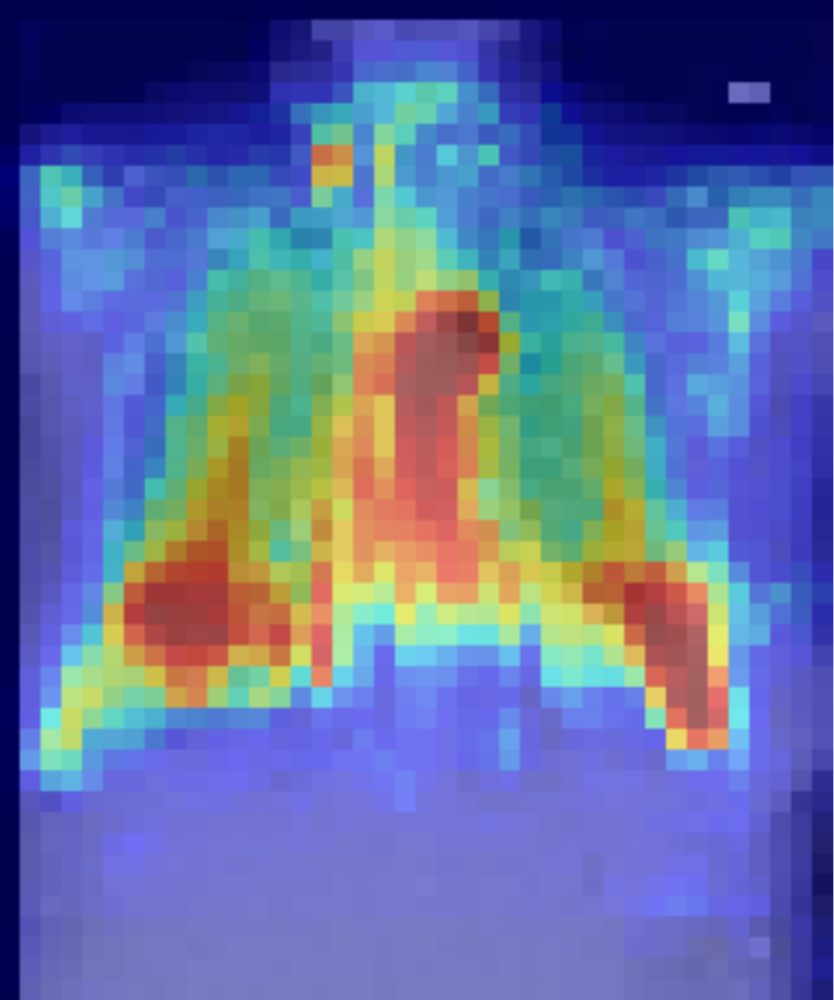}
\includegraphics[width = \figwidth\textwidth]{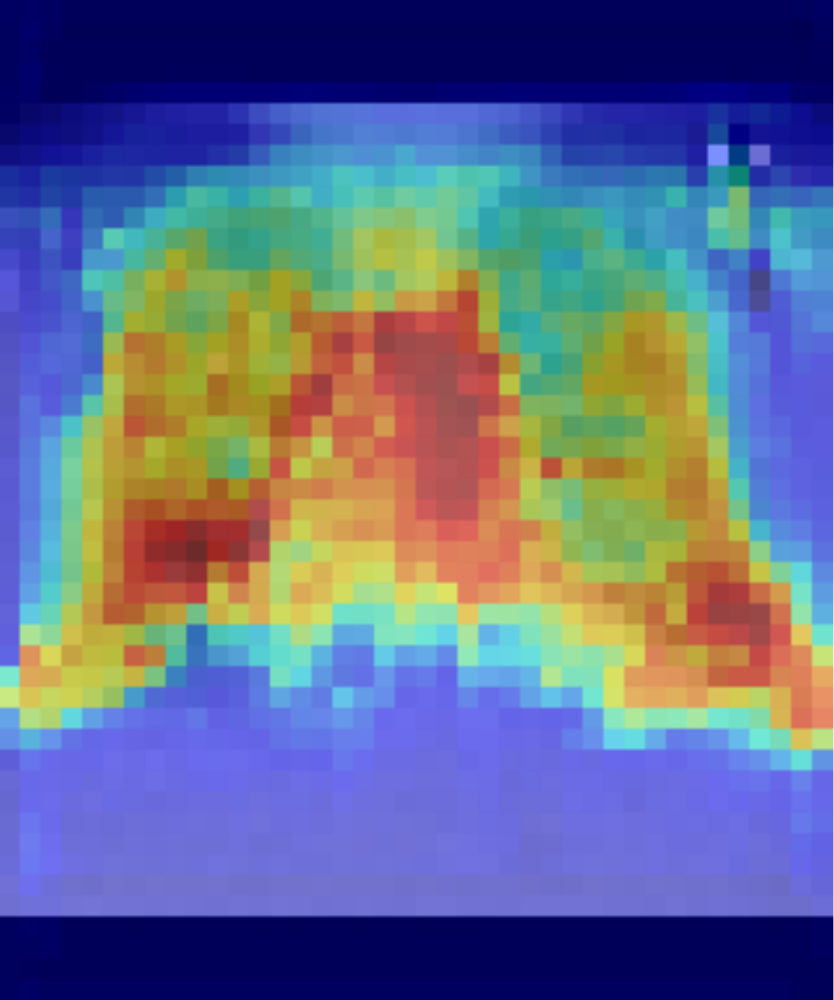}
& 
\includegraphics[width = \figwidth\textwidth]{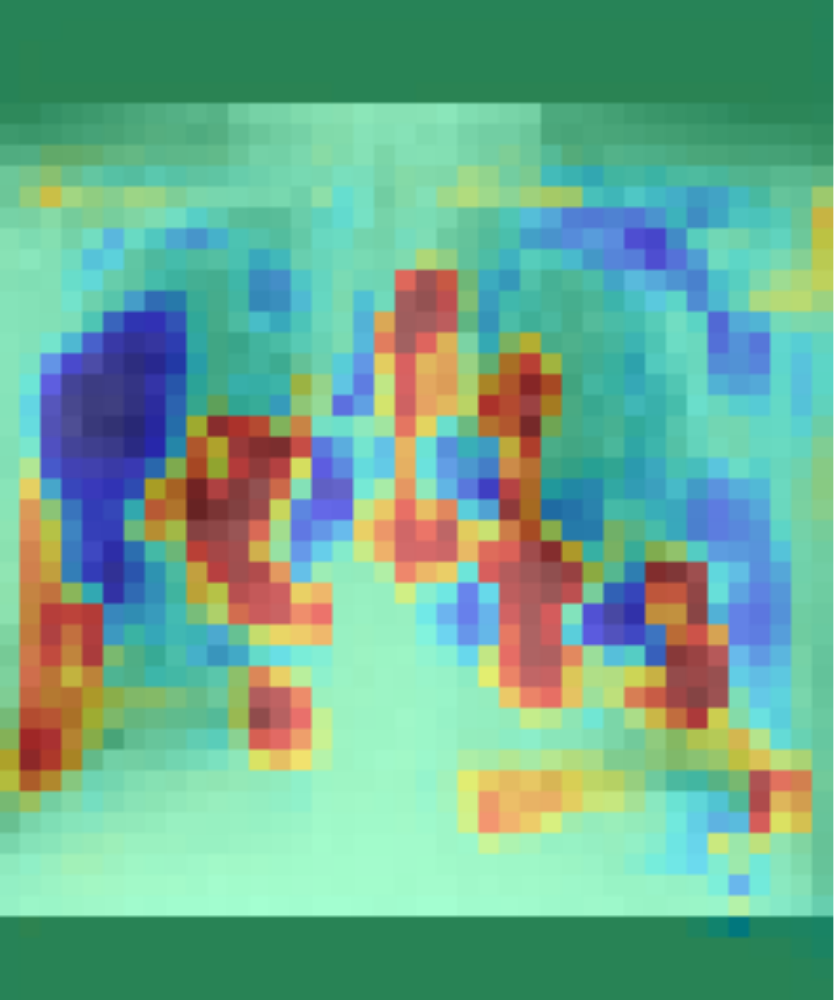}
\includegraphics[width = \figwidth\textwidth]{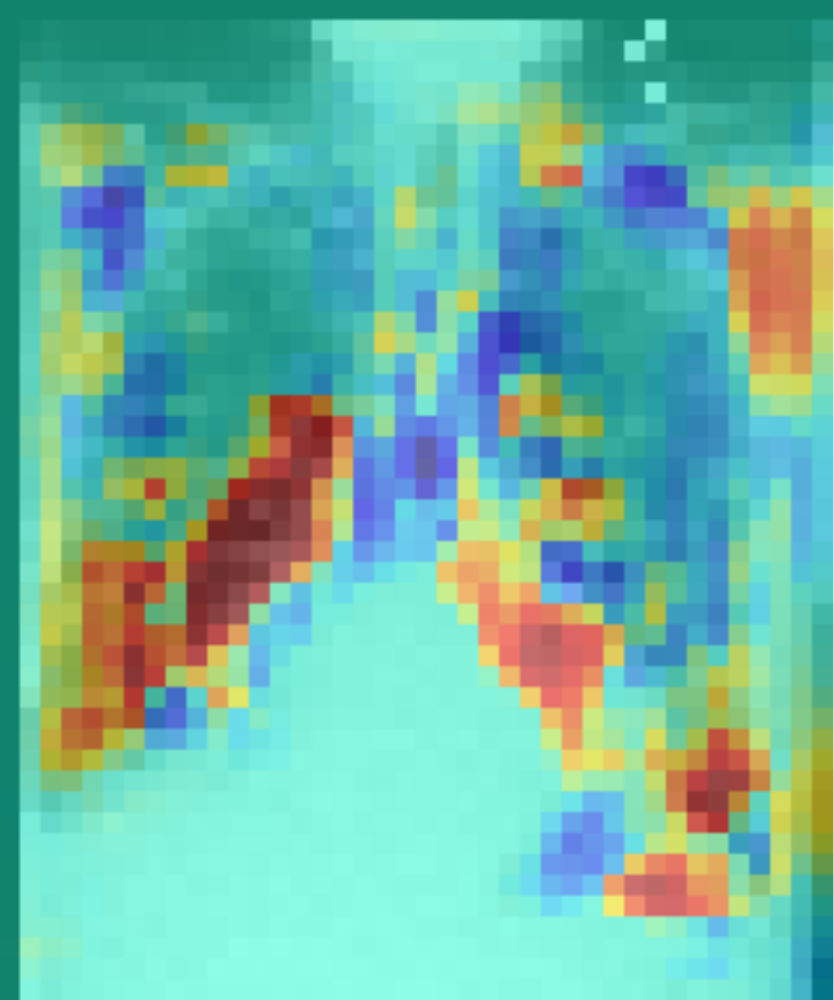}
\includegraphics[width = \figwidth\textwidth]{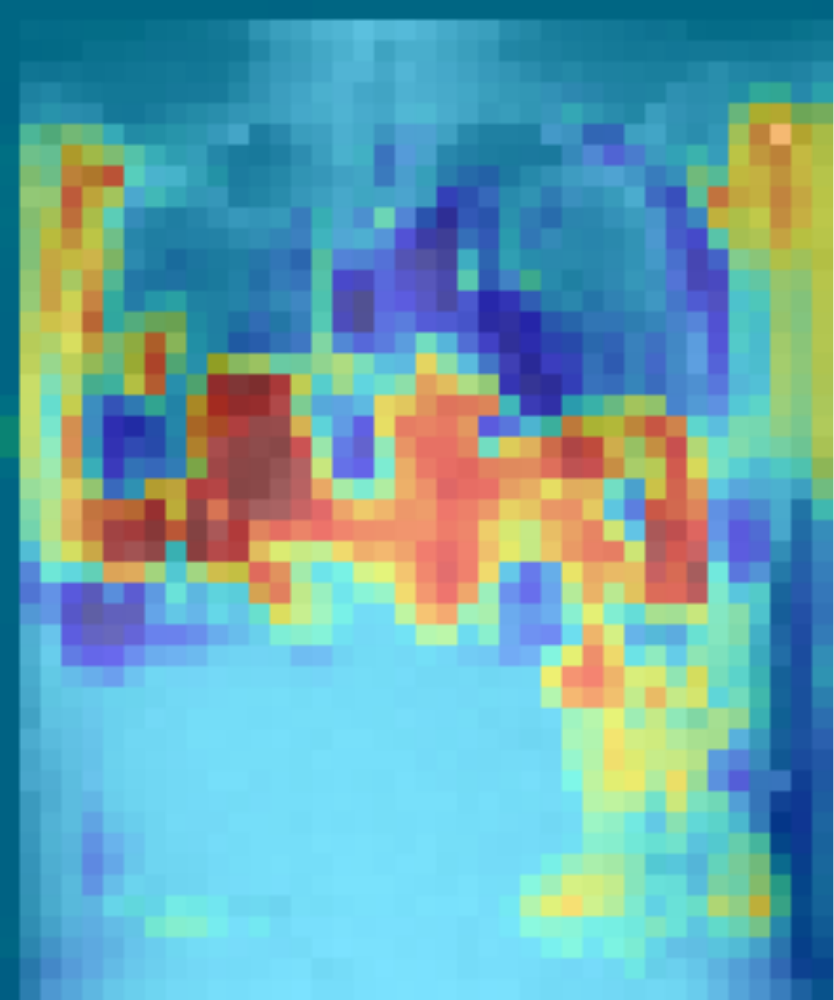}
& 
\includegraphics[width = \figwidth\textwidth]{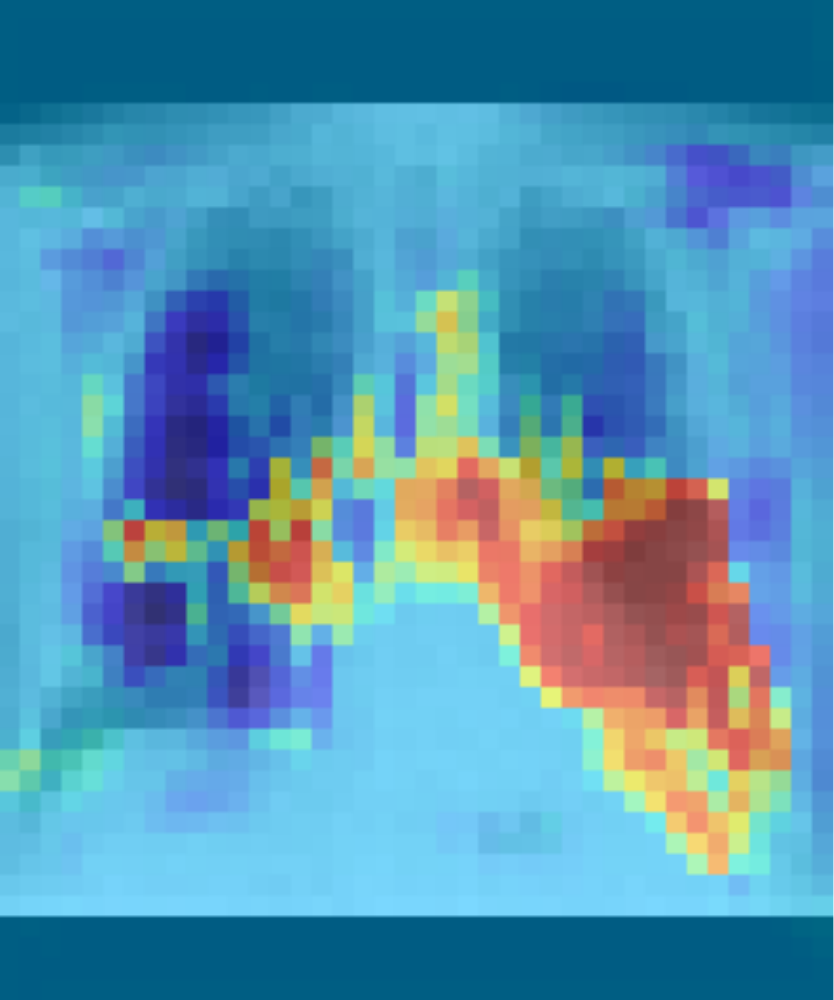}
\includegraphics[width = \figwidth\textwidth]{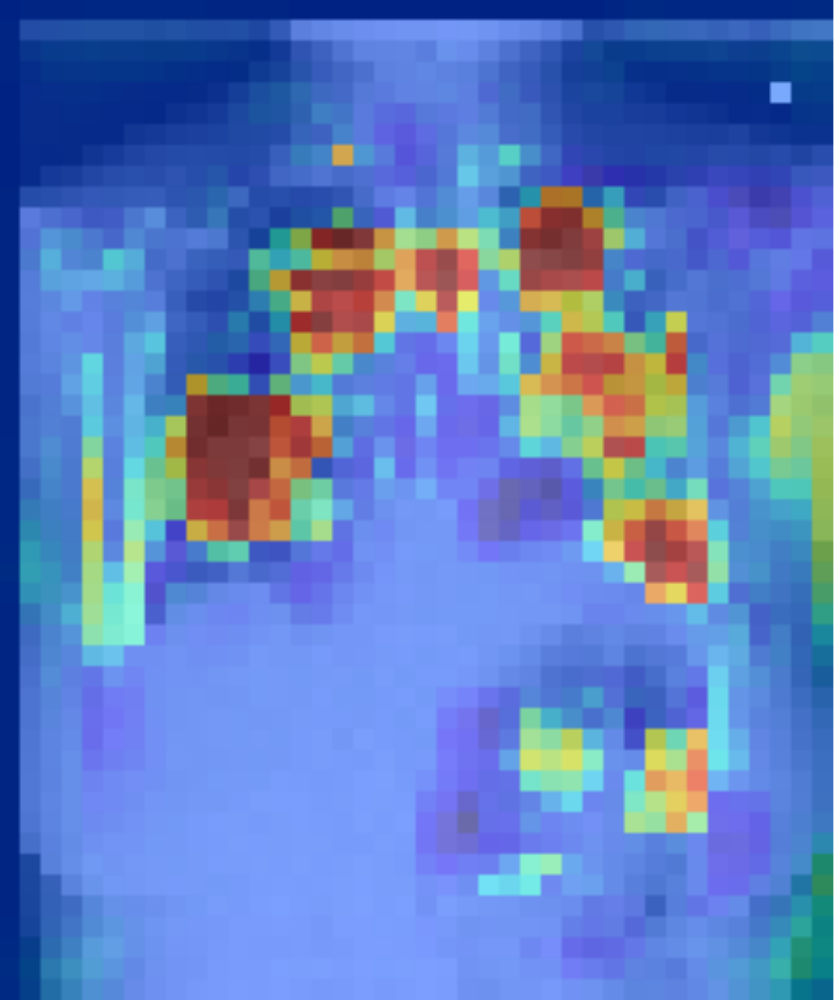}
\includegraphics[width = \figwidth\textwidth]{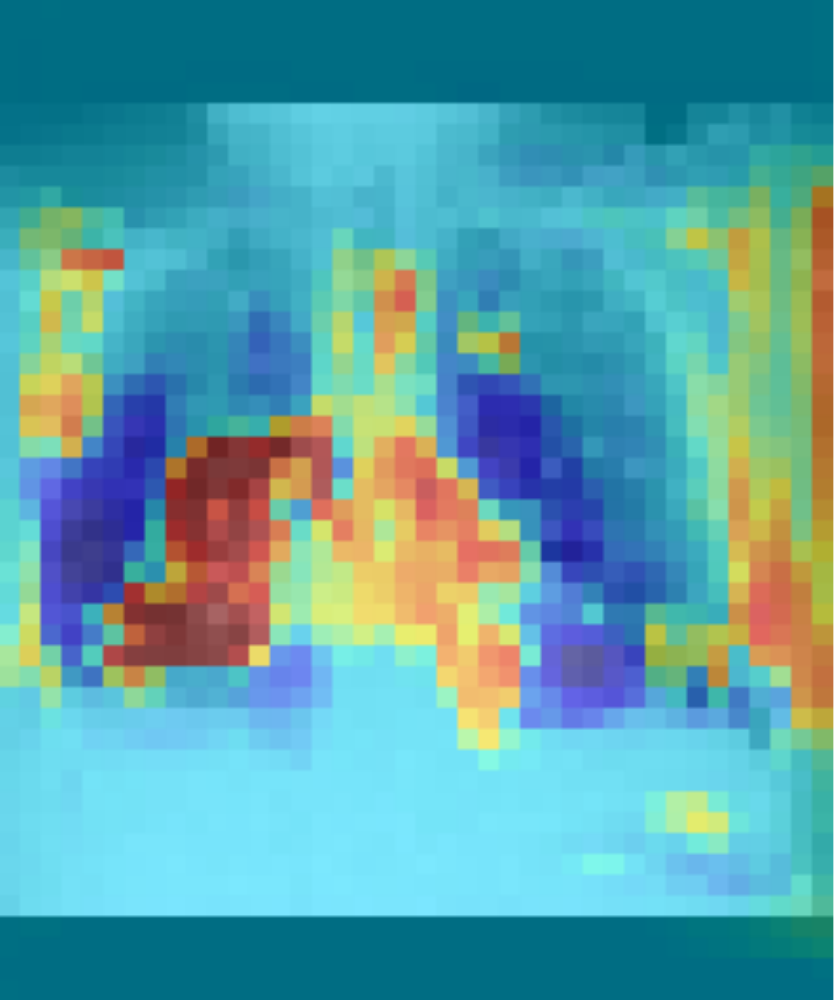}
\\
\begin{turn}{90} EffNet  \end{turn}
&
\includegraphics[width = \figwidth\textwidth]{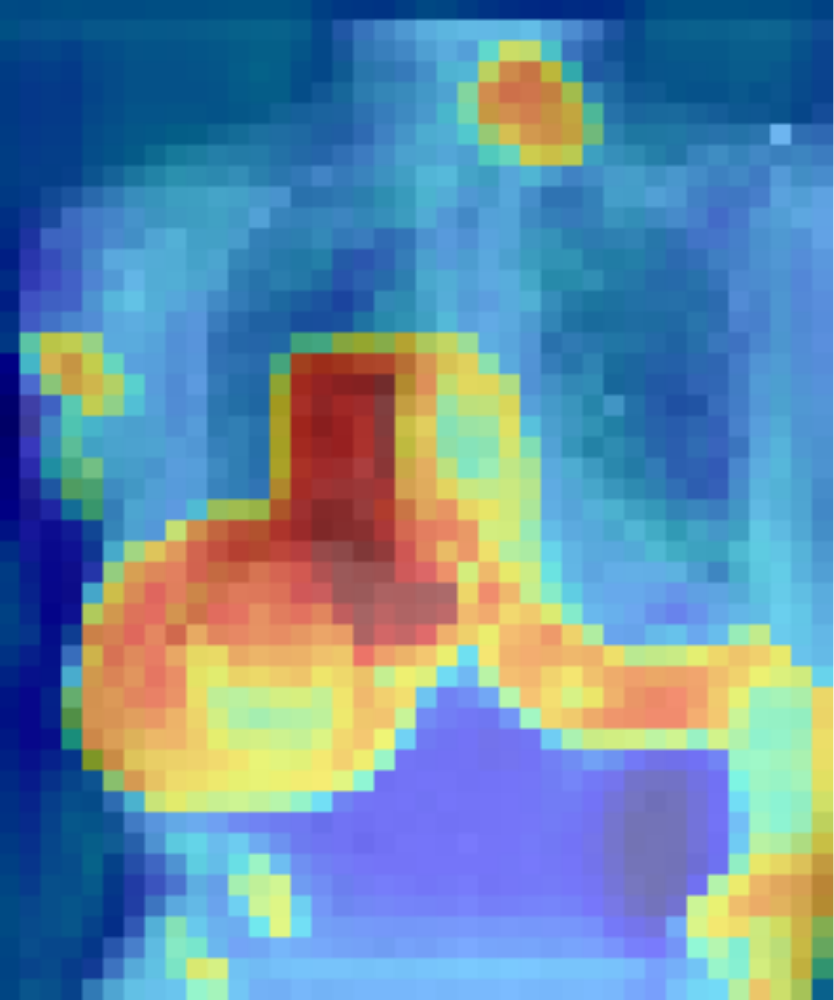}
\includegraphics[width = \figwidth\textwidth]{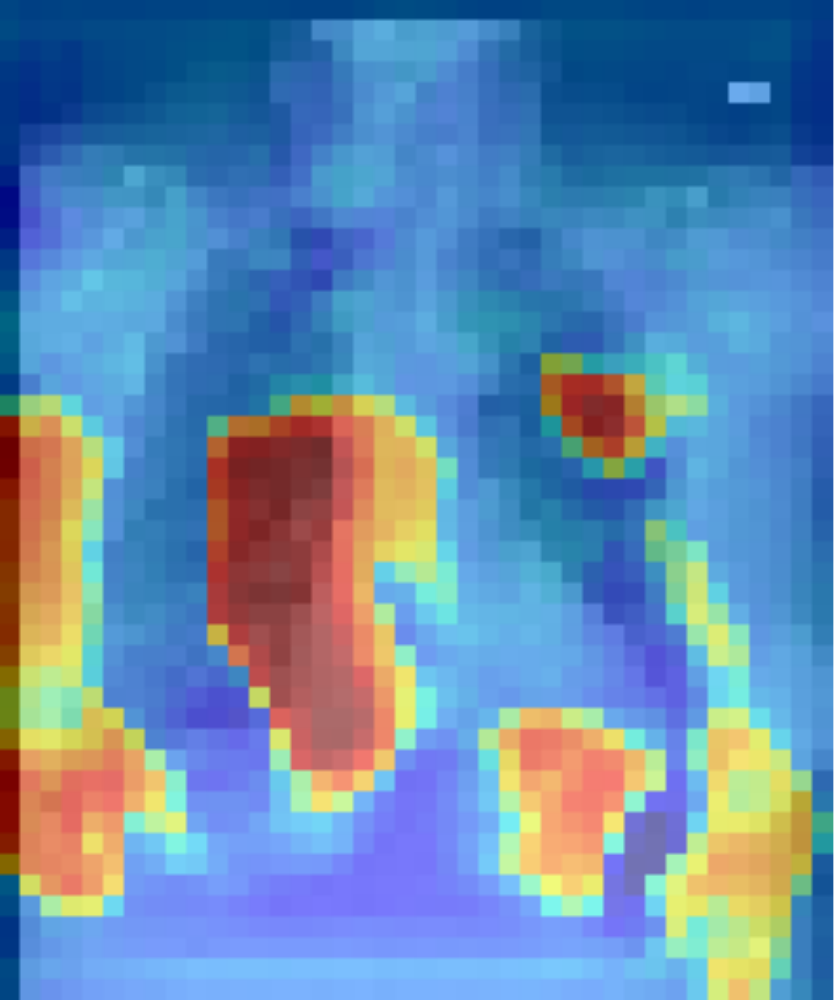}
\includegraphics[width = \figwidth\textwidth]{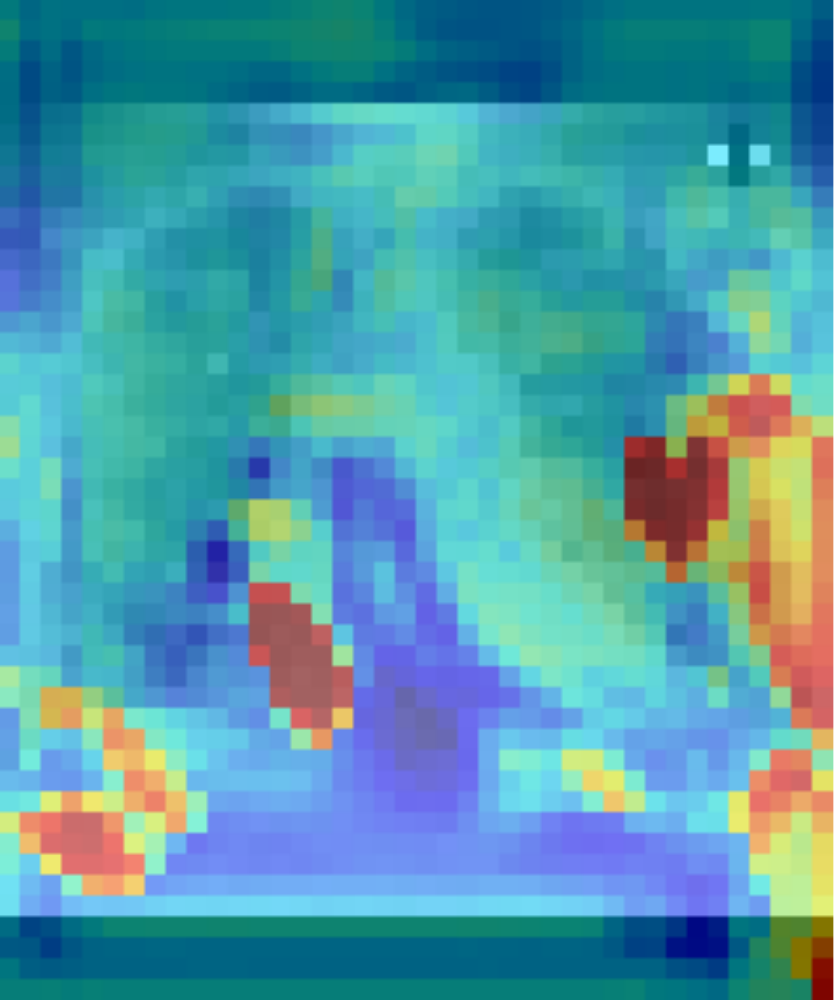}
& 
\includegraphics[width = \figwidth\textwidth]{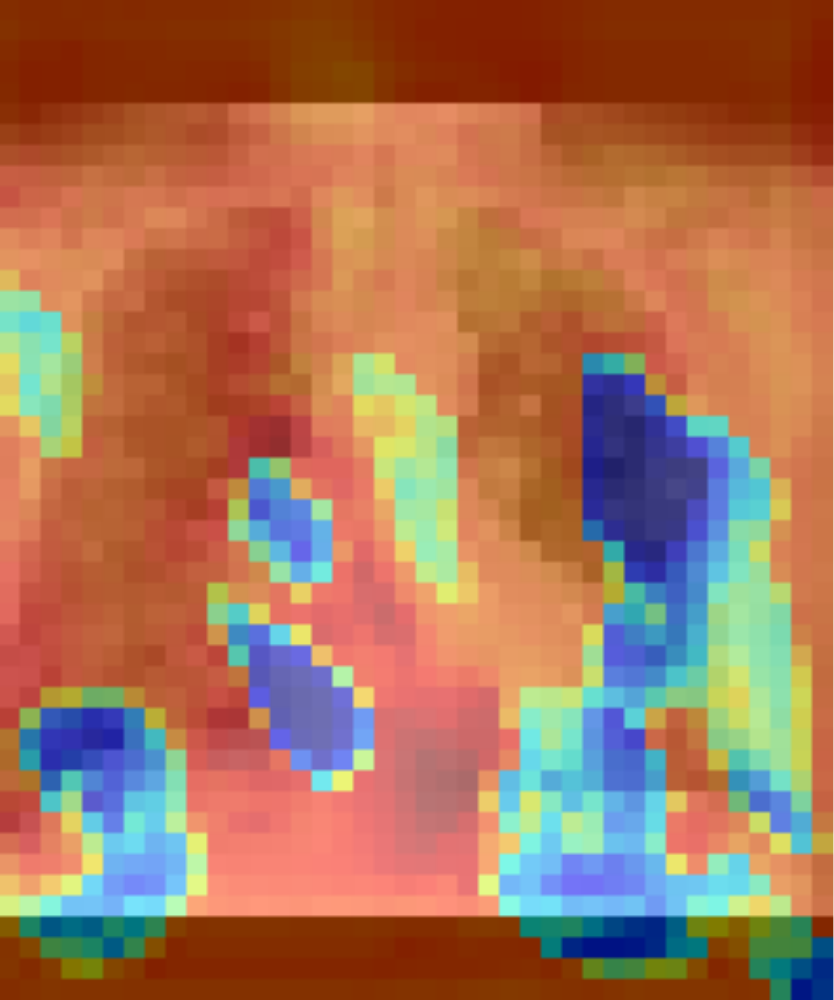}
\includegraphics[width = \figwidth\textwidth]{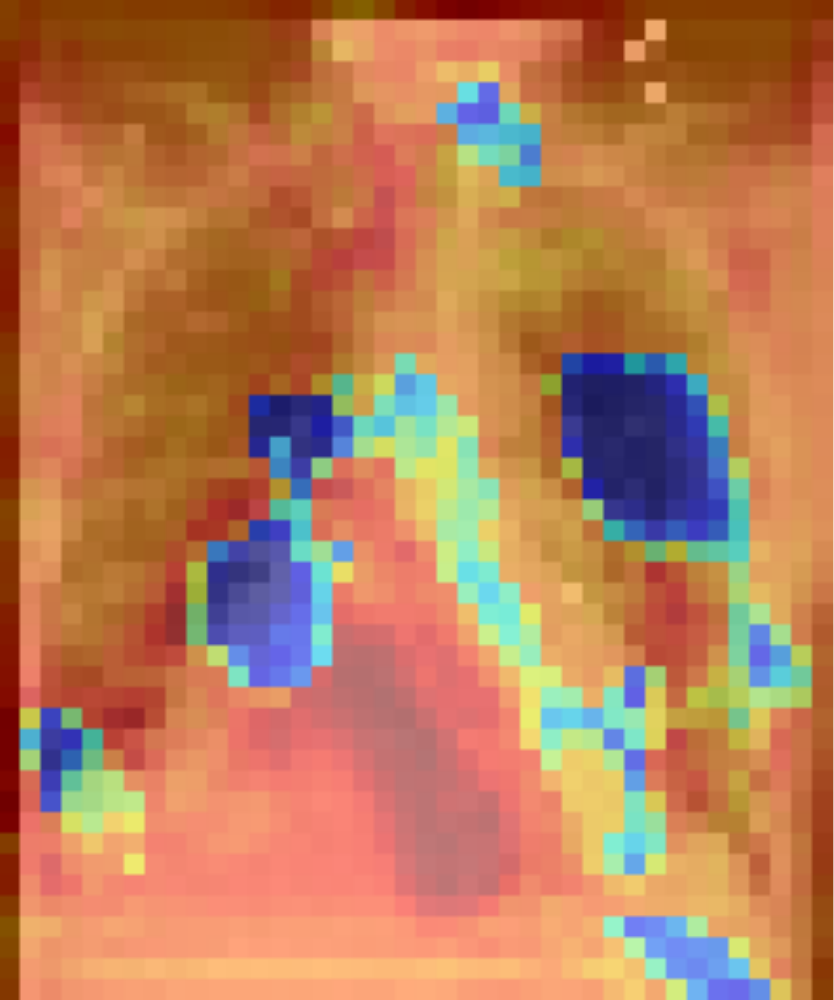}
\includegraphics[width = \figwidth\textwidth]{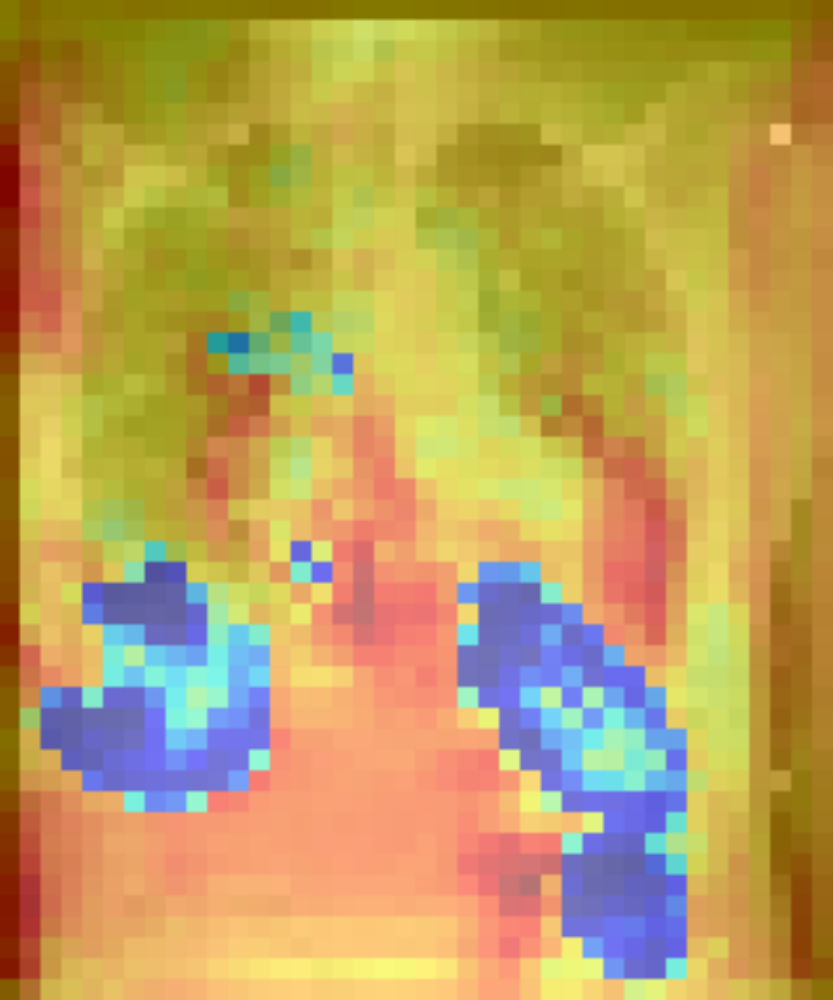}
& 
\includegraphics[width = \figwidth\textwidth]{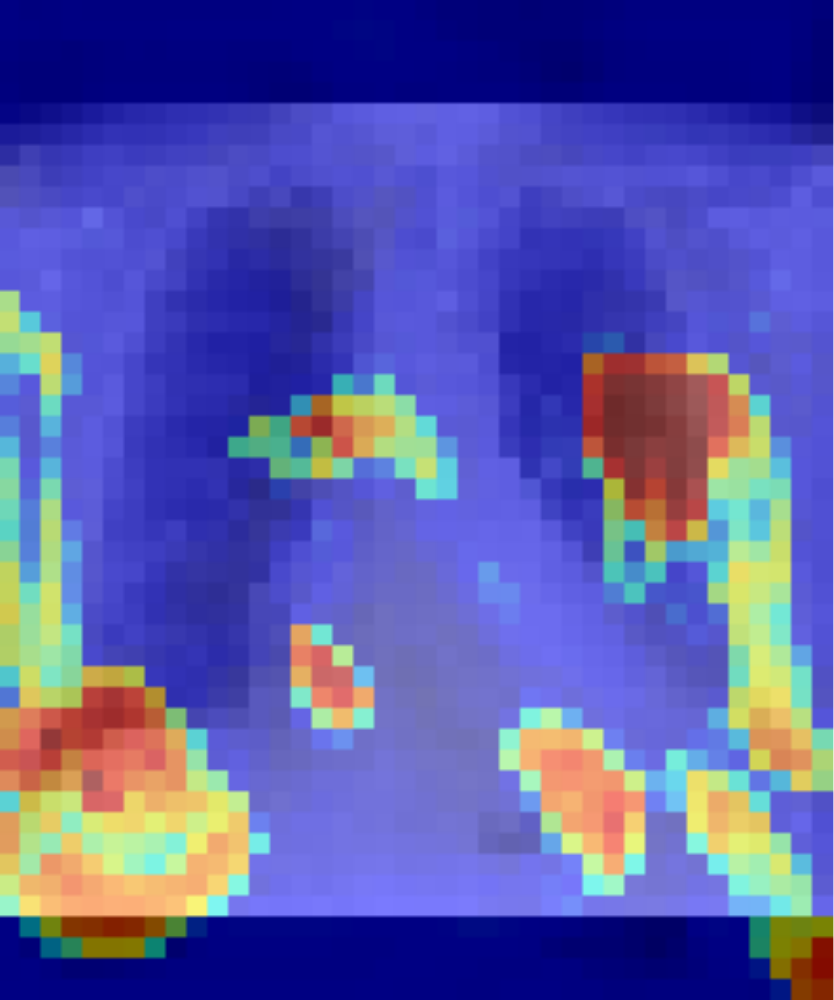}
\includegraphics[width = \figwidth\textwidth]{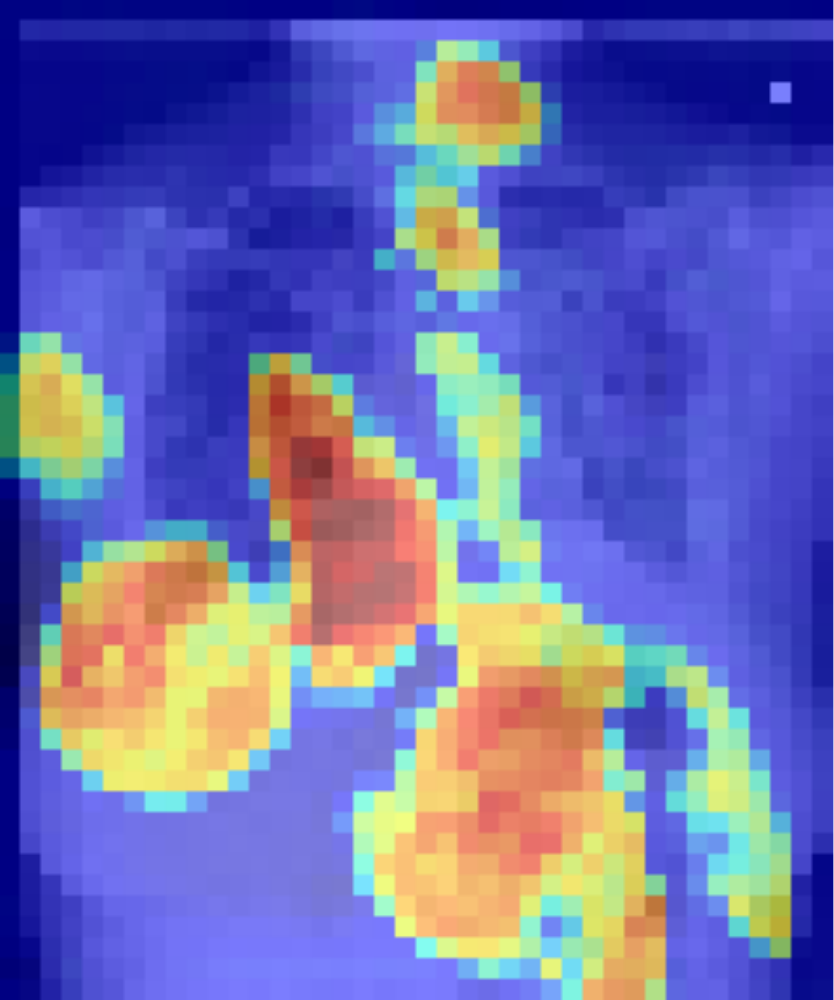}
\includegraphics[width = \figwidth\textwidth]{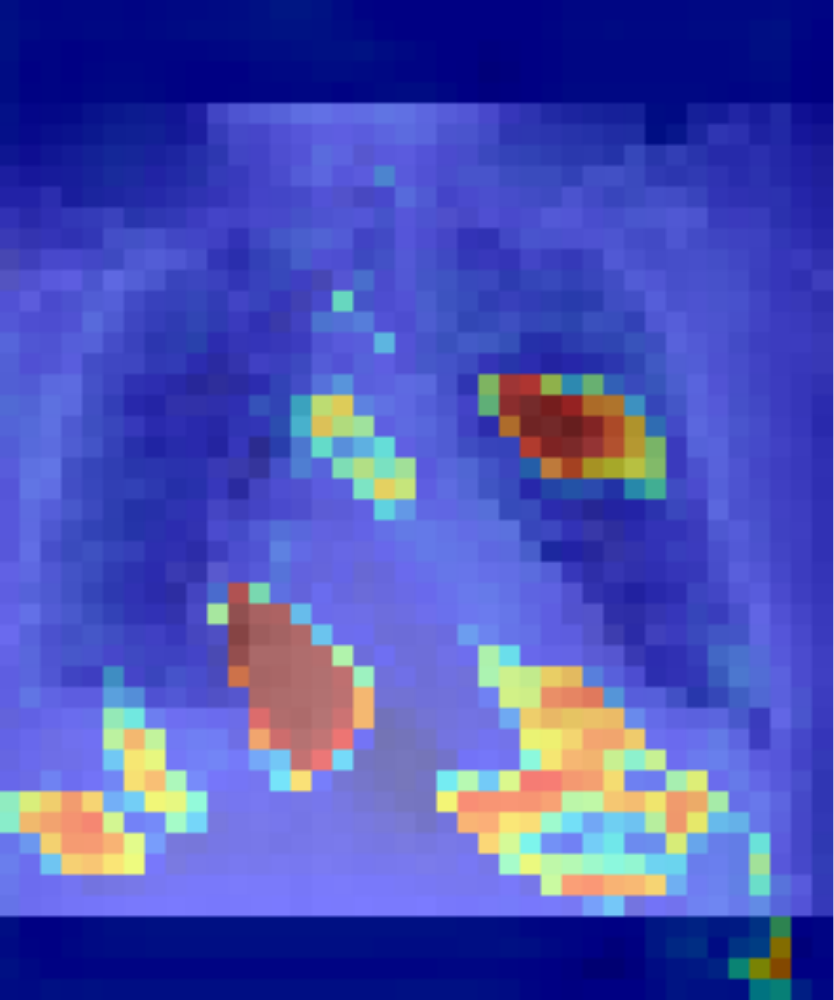}
\\
\begin{turn}{90} ResNet  \end{turn}
&
\includegraphics[width = \figwidth\textwidth]{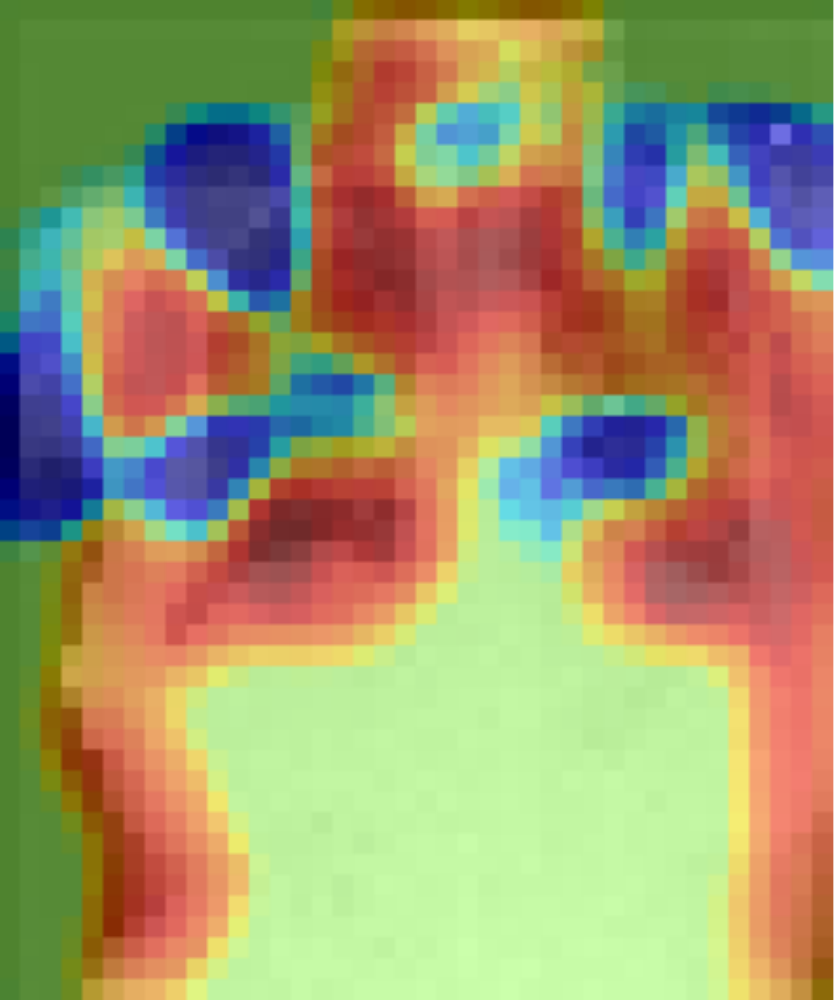}
\includegraphics[width = \figwidth\textwidth]{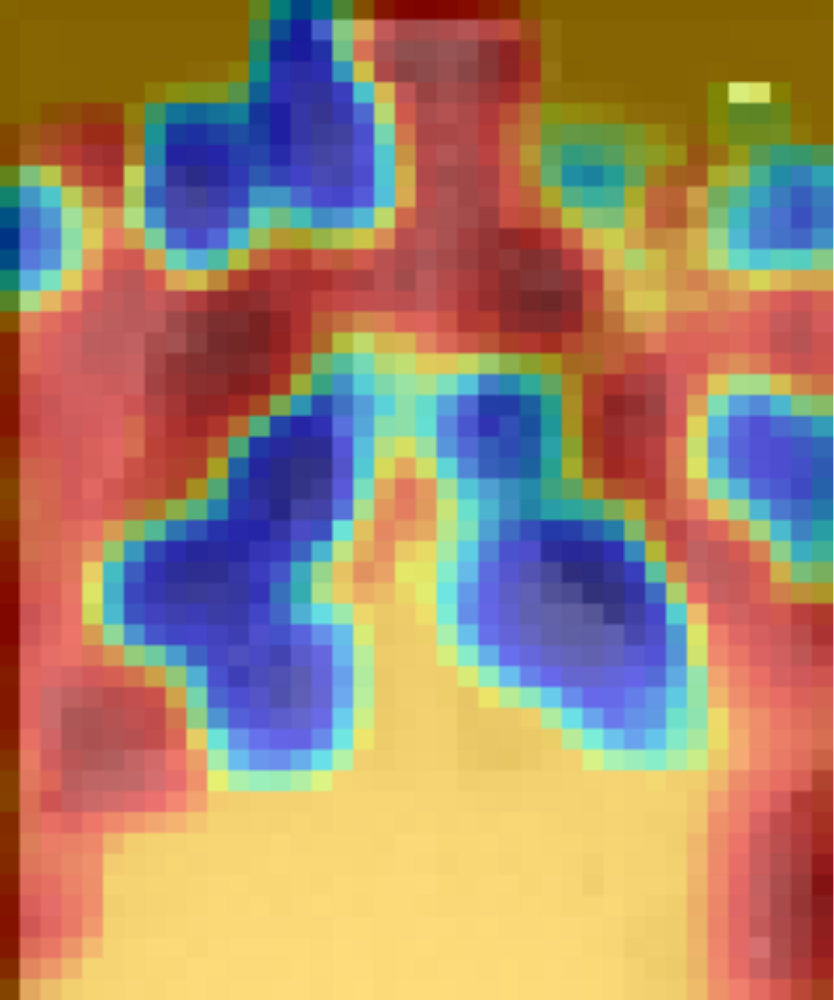}
\includegraphics[width = \figwidth\textwidth]{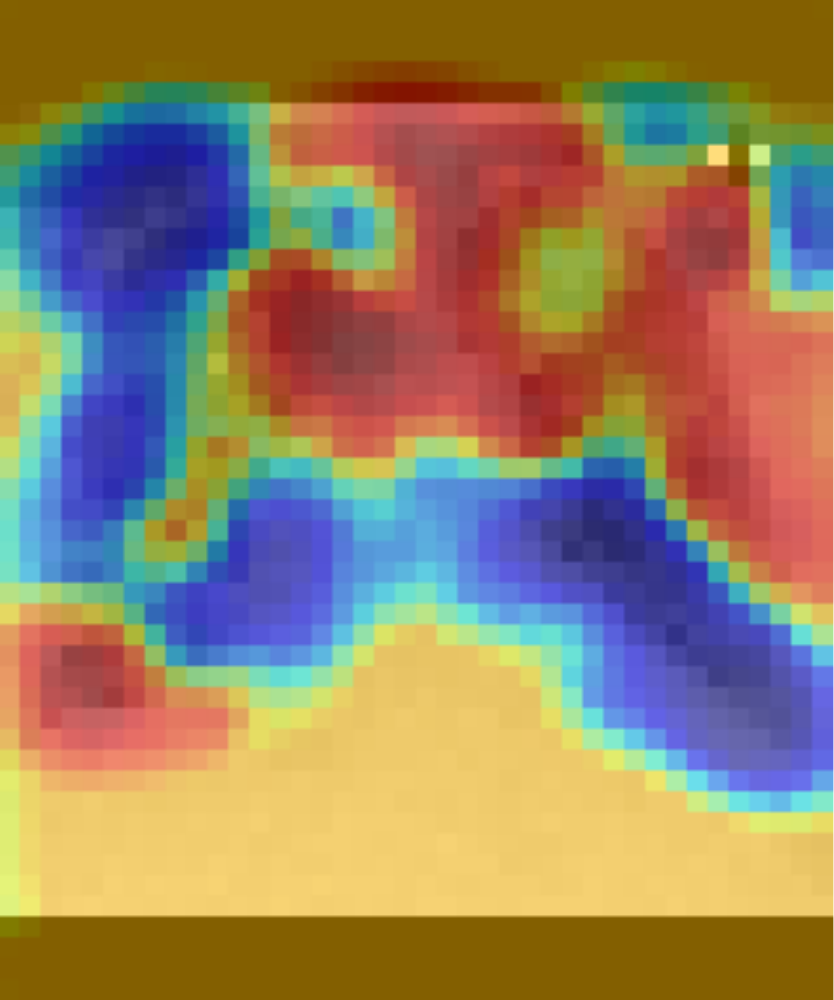}
& 
\includegraphics[width = \figwidth\textwidth]{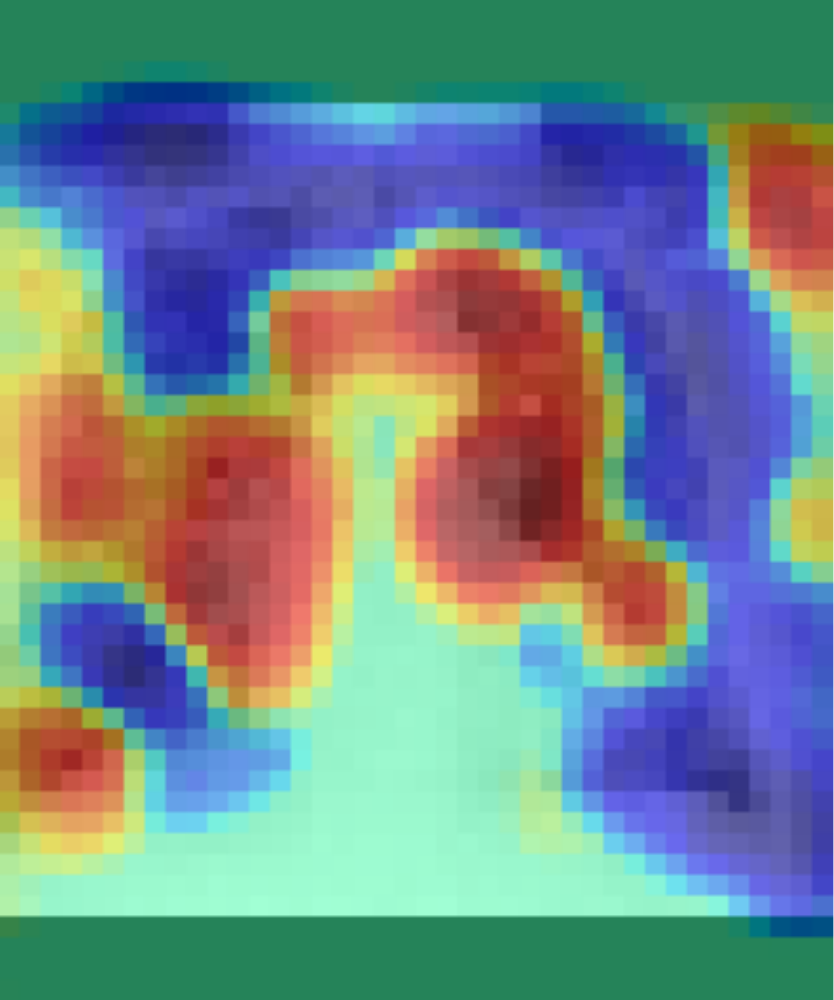}
\includegraphics[width = \figwidth\textwidth]{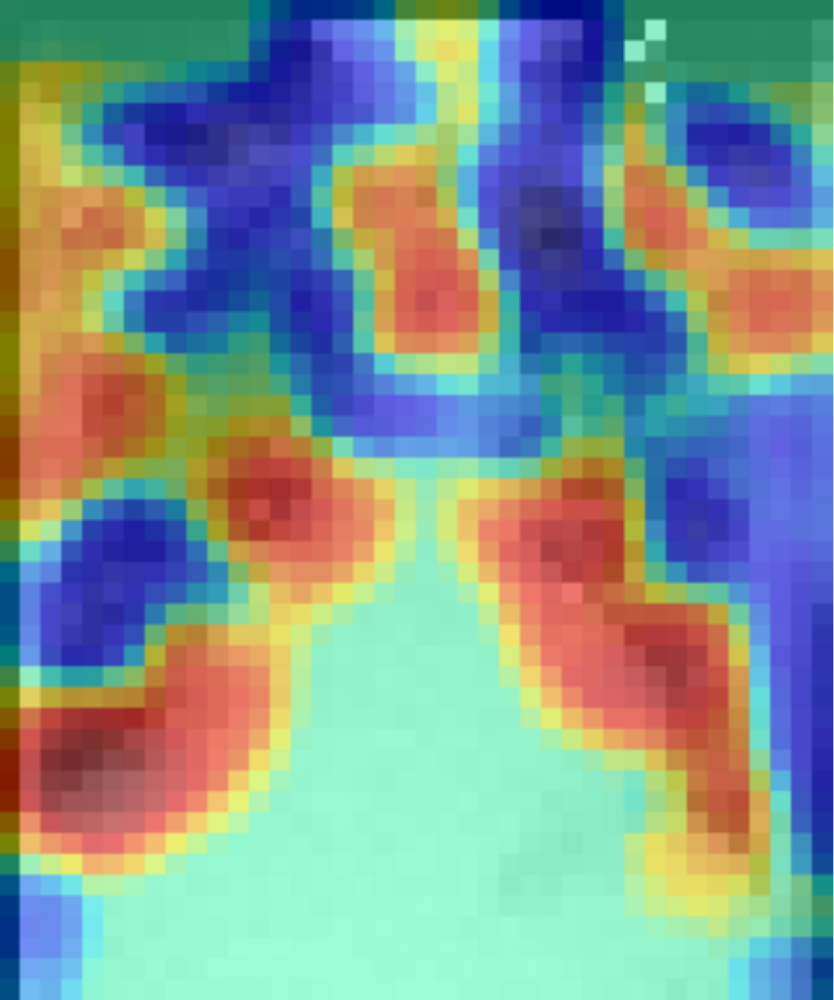}
\includegraphics[width = \figwidth\textwidth]{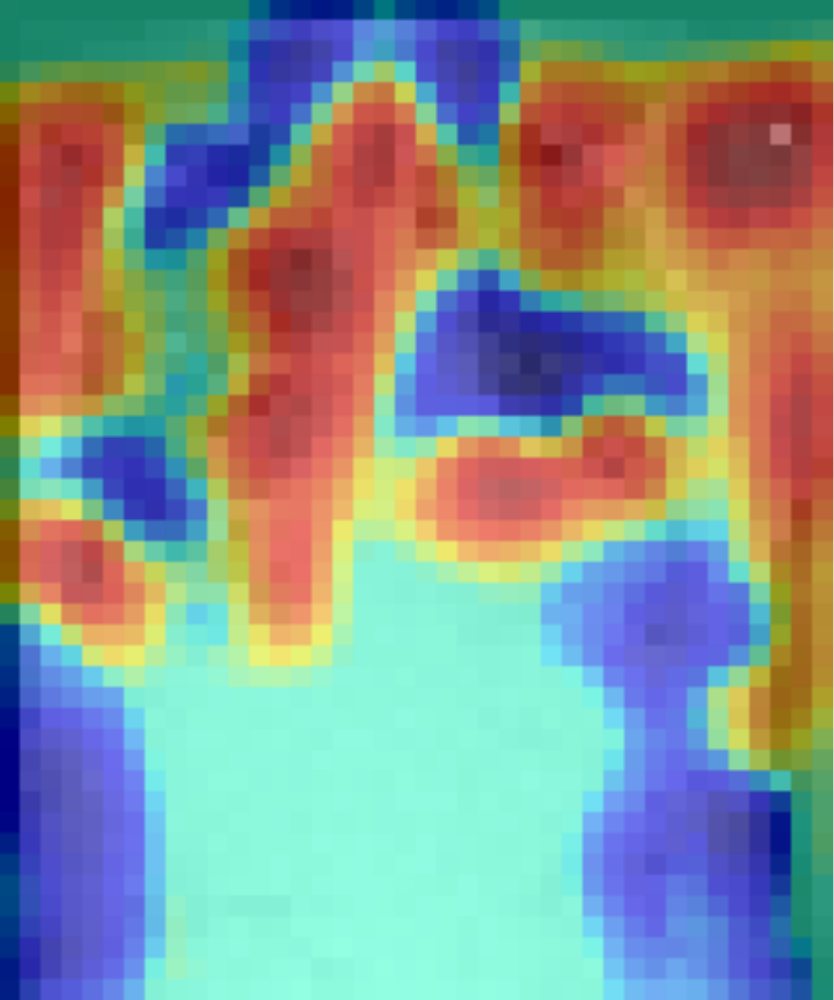}
& 
\includegraphics[width = \figwidth\textwidth]{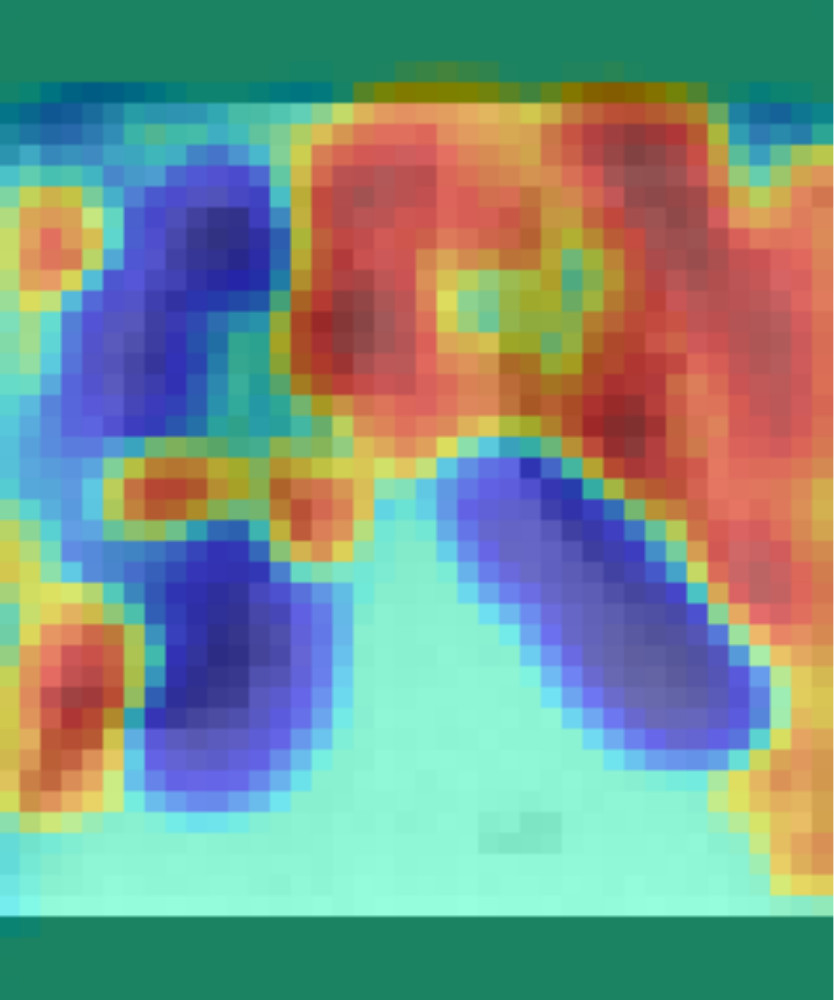}
\includegraphics[width = \figwidth\textwidth]{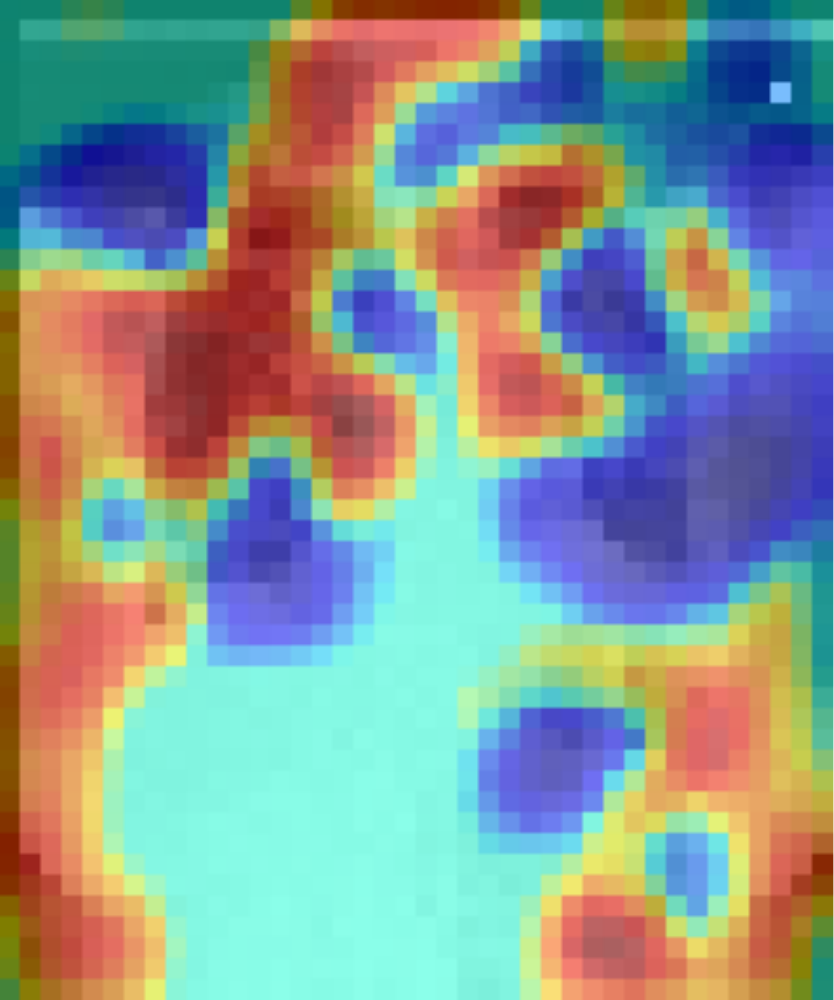}
\includegraphics[width = \figwidth\textwidth]{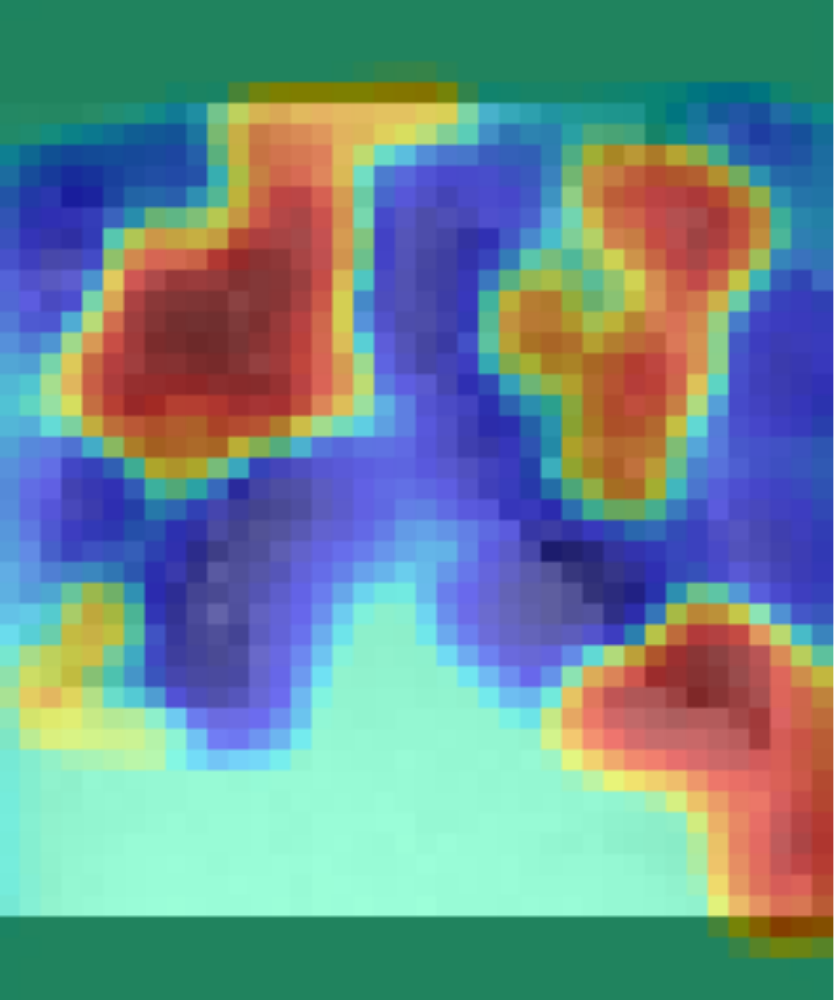}
\\
\bottomrule
\end{tabular}
\caption{Visualization of $\bm\Psi$ and $\bm\Omega^Y$ ($Y=1,2,3$) for each CNNs. The heat maps are plotted on top of the CXR image for better visualization. The heat maps are color coded where red/orange indicates high concentration of attention and blue/green indicates low attention distribution.}
\label{tab:cam visual}
\end{table*}

\subsection{Classification performance}
To evaluate the networks' classification performance, we use the multi-class area under curve (AUC) metrics \cite{hand2001simple}.
Comparative results are shown in Figures \ref{fig:AUC_ENet} and \ref{fig:AUC_RNet}. We use the ANOVA test to check if the mean AUC value for standard CNN is statistically different from that in the GG-CAM modified CNN. The $p$-values for the ANOVA tests are shown at the top of each figure. 
A $p$-value smaller than $0.05$ indicates the difference is statistically significant. Therefore, we conclude that GG-CAM modified CNN can improve the classification significantly as compared to the corresponding standard CNN. More specifically, from $0.723$ to $0.801$, the median AUC for EffNet+GG-CAM is $0.078$ greater than that for EffNet. From $0.721$ to $0.776$, the median AUC for ResNet+GG-CAM is $0.056$ greater than that for ResNet. We focus on median statistics in this paper to avoid influences from outliers, which may occur occasionally (less than $5\%$) when the networks fail to converge. 
In Table \ref{tab:metrics}, we report more detailed classification performance on each label. We can see that the difficulty in classifying pneumonia is greater than for other labels. The performance of EfficientNetv2 is superior to ResNet, and GG-CAM modification generally improves the network's performance. 
\par
We also tested the performance of the network in \cite{wu2020comparison} and \cite{wong2020robust}, termed as PNet, with our dataset. The results suggests that the AUC metrics from PNet are statistically equivalent to that in ResNet and EffNet according to the ANOVA test ($p$>$0.05$). Given that the number of trainable parameters in PNet is at least twice as large as in ResNet or EffNet, and is a customized hybrid model that also uses ResNet in its structure, we did not further apply the GG-CAM modification to PNet for analysis.

\subsection{Interpretability}
We know that the pathologies that we try to classify are organ specific: cardiomegaly should only occur at a patient's heart and pneumonia is a lung infection. Therefore, an interpretable classifier with an abnormal CXR image would put attention to the corresponding organ/anatomical areas on the image. Based on this rationale, we can apply the evaluation method in \cite{selvaraju2017grad}, where the authors calculate the percentage of CAM attention heat maps whose peak lies within the region of the corresponding classification objectives. A larger percentage indicates that the network is more likely to put its attention to the legitimate areas on the input image, and thus more interpretable. 
Mathematically, in our dataset, for an input $\bm{I}$ with true label $Y$ (cardiomegaly or pneumonia), we declare the network's attention is interpretable if:
\begin{equation}
    \underset{i,j}{\operatorname{argmax}}\left(\bm\Omega^Y_{i,j}\right)\in\bm{S}^Y
\end{equation}
where $\bm{S}^2$ is the set of image coordinates for heart segmentation, and $\bm{S}^3$ is the set of image coordinates for lung segmentation. 
Note that the dimension of $\bm\Omega^Y$ is different from that of the input $\bm{I}$. 
Therefore, coordinates in $\bm{S}^Y$ are scaled to match the dimension of $\bm\Omega^Y$. 
Hence, we can calculate the interpretability of the network for labels cardiomegaly and pneumonia, and Figures \ref{fig:heart_ENet}, \ref{fig:heart_RNet}, \ref{fig:lung_ENet}, and \ref{fig:lung_RNet} show the result. 
We also applied the ANOVA test for comparative analysis.
We can see that for cardiomegaly, the variance of the interpretability is large for all CNNs. A statistically significant improvement for interpretability with GG-CAM modification only occurs for ResNet but not for EffNet. For pneumonia, the GG-CAM modification not only improves the interpretability but also reduces the variance of the interpretability distribution.
\par
To visualize network attention and visual attention, in Table \ref{tab:cam visual}, we present $\bm\Omega^Y$ and $\bm\Psi$ for each CNN with the testing images.
By observing the figures, we can see that network attention, as compared to human visual attention, is different. Concentrated red areas of the visual heat map indicates that the expert's attention is specific to the targeted regions, which usually are the abnormal areas. Whereas the network attention is much less concentrated and covers a broader extension of spaces.
Despite the differences, patterns of attention from GG-CAM modified CNNs can be observed. For $Y=1$ (normal CXR images), attention for EffNet+GG-CAM and ResNet+GG-CAM is allocated to the entire heart and lung regions in the image, which can be interpreted as the network checking if any abnormality is present in these regions. For $Y=2$ (cardiomegaly CXR images), heart and neighbouring regions are attended. For $Y=3$ (pneumonia CXR images), the network attention approximately follow the visual attention, highlighting potentially infected lung sections.
However, the attention for EffNet(standard) and ResNet(standard) is more random and less interpretable.

\subsection{Discussion}
Though the GG-CAM modified CNNs usually produce better classification results, disadvantages and limitations awaiting future improvements still exist. The first is the extended training time. Experimental result shows that a GG-CAM modified CNN often requires more epochs ($200$ to $300$ epochs) to converge as compared to the standard CNN (usually less than $150$ epochs). Secondly, the GG-CAM is only applicable to networks with architectures shown in Figure \ref{fig:CAM-CNN}. As human attention is not limited to the application of classification, it may be worthwhile to extend the GG-CAM method to a broader range of networks and/or tasks. 
Finally, we have only validated our method in a single radiology task, which follows well-established examination standards and protocols. However, as human attention exhibits diversified patterns across individuals, it is worth exploring how our method works when the tasks do not have structured standards and protocols. 

\section{Conclusion}\label{sec:Conclusion}
In this paper, we showed that human attention can be used to directly regulate the network attention that not only boosts the network's performance but also enhances the interpretability. With GG-CAM, most classification CNNs can be modified to gain the capability of augmenting their own attention with human attention. Therefore, by recording the gaze behavior, we can develop better methods to alleviate the workload for medical practitioners and specialists performing time-consuming and labour-intensive CXR annotation tasks. 
\par
While we are still at an stage of understanding artificial and biological neural networks, we believe our results can provide insights for future interdisciplinary research and applications in attention and computer vision.

{\small
\bibliographystyle{ieee_fullname}
\bibliography{egpaper.bib}
}

\end{document}